# Foreground separation and constraints on primordial gravitational waves with the PICO space mission

Ragnhild Aurlien,[a] Mathieu Remazeilles,[b] Sebastian Belkner,[c] Julien Carron,[c] Jacques Delabrouille,[d] Hans Kristian Eriksen,[a] Raphael Flauger,[e] Unni Fuskeland,[a] Mathew Galloway,[a] Krzysztof M. Górski,[f,g] Shaul Hanany,[h,*] Brandon S. Hensley,[i] J. Colin Hill,[j,k] Charles R. Lawrence,[f] Clement Pryke,[h] Alexander van Engelen[l] and Ingunn Kathrine Wehus[a]

[a]Institute of Theoretical Astrophysics, University of Oslo, Blindern, Oslo, Norway
[b]Instituto de Fisica de Cantabria (CSIC-UC),
 Avda. los Castros s/n, 39005 Santander, Spain
[c]Université de Genève, Département de Physique Théorique et CAP,
 Genève 4, Switzerland
[d]Centre Pierre Binétruy International Research Laboratory, CNRS,
 UC Berkeley and LBNL, Berkeley, CA 94720, U.S.A.
[e]UC San Diego, La Jolla, CA, 92093, U.S.A.
[f]Jet Propulsion Laboratory, California Institute of Technology,
 4800 Oak Grove Drive, Pasadena, CA 91109, U.S.A.
[g]Warsaw University Observatory, Aleje Ujazdowskie 4, 00-478 Warszawa, Poland
[h]University of Minnesota — Twin Cities, 115 Union St. SE, Minneapolis, MN, 55455, U.S.A.
[i]Department of Astrophysical Sciences, Princeton University, Princeton, NJ 08544, U.S.A.
[j]Department of Physics, Columbia University, New York, NY 10027, U.S.A.
[k]Center for Computational Astrophysics, Flatiron Institute,
 New York, NY 10010, U.S.A.
[l]School of Earth and Space Exploration, Arizona State University, Tempe, AZ 85287, U.S.A.

E-mail: hanany@umn.edu

*Corresponding author.

https://doi.org/10.1088/1475-7516/2023/06/034

**Abstract.** PICO is a concept for a NASA probe-scale mission aiming to detect or constrain the tensor to scalar ratio $r$, a parameter that quantifies the amplitude of inflationary gravity waves. We carry out map-based component separation on simulations with five foreground models and input $r$ values $r_{\rm in} = 0$ and $r_{\rm in} = 0.003$. We forecast $r$ determinations using a Gaussian likelihood assuming either no delensing or a residual lensing factor $A_{\rm lens} = 27\%$. By implementing the first full-sky, post component-separation, map-domain delensing, we show that PICO should be able to achieve $A_{\rm lens} = 22\% - 24\%$. For four of the five foreground models we find that PICO would be able to set the constraints $r < 1.3 \times 10^{-4}$ to $r < 2.7 \times 10^{-4}$ (95%) if $r_{\rm in} = 0$, the strongest constraints of any foreseeable instrument. For these models, $r = 0.003$ is recovered with confidence levels between $18\sigma$ and $27\sigma$. We find weaker, and in some cases significantly biased, upper limits when removing few low or high frequency bands. The fifth model gives a $3\sigma$ detection when $r_{\rm in} = 0$ and a $3\sigma$ bias with $r_{\rm in} = 0.003$. However, by correlating $r$ determinations from many small 2.5% sky areas with the mission's 555 GHz data we identify and mitigate the bias. This analysis underscores the importance of large sky coverage. We show that when only low multipoles $\ell \leq 12$ are used, the non-Gaussian shape of the true likelihood gives uncertainties that are on average 30% larger than a Gaussian approximation.

# Contents

## 1 Introduction

Observations of the cosmic microwave background (CMB) from three space missions, COBE-DMR, WMAP, and Planck have established ΛCDM as the widely accepted standard model of cosmology, have given a wealth of information about the evolution of structures in the Universe, and have been used to support numerous astrophysical investigations. Several missions have been proposed to take the next leap in understanding the evolution of the Universe, including LiteBIRD [1], which has been approved by the Japanese space agency JAXA, CORE [2], CMB-Bhārat [3], and PICO [4]. PICO, The Probe of Inflation and Cosmic

Origins, a concept for a NASA-led probe-scale space mission [4], is the most sensitive of the proposed next generation space missions. For example, relative to LiteBIRD it would give maps that are deeper by at least a factor of 2.5 with nearly four times higher angular resolution in overlapping bands. PICO is expected to give the tightest constraints on cosmological parameters compared to all forthcoming or planned CMB instruments.

PICO has seven science objectives, five of which are to be extracted from the highest ever signal-to-noise ratio (S/N) full-sky maps of the CMB. After PICO's prime mission of 5 years, the combined map-noise level over the entire sky would be 0.6 $\mu$K arcmin. Since the mission has no liquid cryogens, longer lifetime and lower noise are likely. Of the numerous probe-scale mission concepts submitted to the Astro2020 panel, a next-generation CMB mission was one of three recommended for further development this decade for a possible flight in the 2030s [5].

PICO's most demanding requirement is the level of constraint on the tensor to scalar ratio $r$, which quantifies the amplitude of gravitational waves produced during the epoch of inflation shortly after the big bang. For the case of a null detection, the requirement is to achieve $r \leq 0.0002 \, (95\%)$. If $r \neq 0$ the requirement is to achieve $5\sigma$ detection of $r = 0.0005$. A priori, in the context of inflation, the expected value of $r$ can range over many orders of magnitude, including values far too small to be detected. However, the simplest models of inflation that have a single inflaton, and that naturally explain the observed value of the scalar spectral index, only have a single free parameter, the distance in field space over which the potential varies appreciably. In several well-motivated scenarios, this scale shares a common origin with the scale of gravity, typically referred to as the Planck scale. Such models predict values $r \gtrsim 0.001$ [6, 7]. The absence of detection at this level would definitively rule out this particularly well-motivated class of inflationary models [8]. Conversely, an unambiguous detection would definitively establish inflation as the source of primordial perturbations, determine the energy scale at which inflation took place, and would give a first direct probe of quantum gravity. The current upper limit is $r < 0.032 \, (95\%)$ [9, 10].

CMB determinations of $r$ rely on measuring the polarization $Q$ and $U$ Stokes parameters, converting them to E- and B-modes, and forming the corresponding EE and BB angular power spectra. The level of $r$ is linearly proportional to the BB power spectrum, $r = 0.15 \cdot \mathcal{D}^{\rm BB}_{\ell=90}/(0.1\,\mu{\rm K})^2$, where $\mathcal{D}^{\rm BB}_{\ell=90} = \ell(\ell+1)C^{\rm BB}_\ell/2\pi$ is the level at $\ell = 90$. Reaching PICO's constraint on inflationary gravitational waves is demanding — even with its high S/N maps — because the B-mode foreground emission within the Milky Way is known to have amplitudes much larger than the levels of $r$ targeted by PICO [9, 11]. Separating the cosmological signal from the foregrounds, a process commonly called "component separation," requires low-noise multi-frequency observations, or accurate prior knowledge, that are unlikely to be available before the mission flies, because it would take a PICO-type mission to map the sky with the requisite S/N.

Gravitational lensing of CMB photons by large scale structures between the surface of last scattering and our telescopes presents another challenge [12]. Lensing scatters photons off their original paths, distorting slightly the original pattern of the CMB anisotropy. In polarization, the primary effect is the conversion of high $\ell$ E-mode power to lower $\ell$ B-mode power. The sample variance of this lensing-induced B-mode acts as a source of approximately white noise. However, with sufficiently high S/N polarimetric measurements extending to $\ell \gtrsim 1000$, CMB maps can be delensed [13–15] and delensing improves constraints on cosmological parameters. PICO's resolution of few arcmin at the main CMB bands near 200 GHz makes internal delensing — the use of its own data to delens — possible [16, 17], which is a key element in achieving the stringent science goal for $r$.

In this paper we use simulations to study whether data from PICO's 21 frequency bands (section 2) would enable foreground cleaning such that the required level of constraint on $r$ can be achieved. We construct sky maps matching five possible foreground models, all of them broadly consistent with current Planck data (section 3). They span a broad range of complexity and input assumptions. The sky maps, which include CMB signals and noise (section 4), approximately represent maps that would be obtained with PICO. The maps are approximate because we assume a spatial noise distribution that is pixel-independent and scale-free with $\ell$, an appropriate approximation at this stage of mission development. We apply both parametric and blind component separation methods and estimate $r$. Details of the methods and results from each are given in sections 5 and 6. Delensing of the sky maps is handled in two ways, an analytic, power spectrum domain approach, and using an iterative, map domain delensing algorithm (section 7). We discuss and summarize in sections 8 and 9.

Testing the component separation capability of various instruments has been carried out in the past. The recent most relevant reports were those provided in the context of CORE [18], LiteBIRD and PICO [1, 19], CMB-Bharat [3], Simons Observatory [20], and CMB-S4 [21]. This paper is unique because with the PICO data we strive to set unprecedented constraints on cosmological parameters, because we analyze a diverse set of five sky foreground models spanning a realistic range of possibilities, and because this is the first paper to report on map-level iterative delensing as an integral part of the analysis pipeline.

## 2 PICO instrument parameters

The PICO mission concept has a single instrument, an imaging polarimeter, operating in 21 frequency bands between 21 and 799 GHz. The instrument consists of a two mirror telescope with a 1.4 m diameter entrance aperture, which is based on an 'open-Dragone' design [22]. The telescope feeds a focal plane populated with 13,000 bolometers and gives a resolution between 1.1 arcmin at the highest frequency and 38.4 arcmin at the lowest. The bolometers are operated from a bath temperature of 0.1 K. PICO will conduct observations from the L2 Lagrange point with a scan pattern that covers the full sky within 6 months. The prime mission duration is 5 years giving 10 redundant full sky surveys, but because there are no consumables, mission lifetime could extend significantly longer. The mission concept has required and estimated noise levels of 0.87 and 0.61 $\mu$K arcmin, respectively, accounting only for the 5 yr prime mission. In this paper we use the estimated map noise levels, as given in table 1. The simulations that are used in this work began before the concept report was finalized and due to late iterations on the design of the focal plane some of the values in the table differ slightly from the values quoted in the more definitive final report [4]. The differences are minor and do not change the combined map noise level. The frequency bands and their noise levels as given in table 1 were not optimized for component separation. Such optimization is left for future work.

## 3 Foreground models

We consider five models for Galactic foreground emission, as described in the following sections and summarized in table 2. For ease of reference, the models are given the names and short names as shown in the first column. For levels of $r$ relevant for upcoming CMB instruments including PICO, $0.5 \times 10^{-3} \leq r \leq 5 \times 10^{-3}$, the B-mode signal is much smaller than Galactic foregrounds, see figure 1. The figure shows the level of foregrounds expected at three frequency

| Frequency (GHz) | 21 | 25 | 30 | 36 | 43 | 52 | 62 | 75 | 90 | 108 | 129 | 155 | 186 | 223 | 268 | 321 | 385 | 462 | 555 | 666 | 799 |
|---|---|---|---|---|---|---|---|---|---|---|---|---|---|---|---|---|---|---|---|---|---|
| FWHM (arcmin) | 38.4 | 32.0 | 28.3 | 23.6 | 22.2 | 18.4 | 12.8 | 10.7 | 9.5 | 7.9 | 7.4 | 6.2 | 4.3 | 3.6 | 3.2 | 2.6 | 2.5 | 2.1 | 1.5 | 1.3 | 1.1 |
| Noise[†] ($\mu$K-arcmin) | 16.9 | 11.8 | 8.1 | 5.7 | 5.8 | 4.1 | 3.8 | 2.9 | 2.0 | 1.6 | 1.6 | 1.3 | 2.6 | 3.0 | 2.1 | 2.9 | 3.5 | 7.4 | 34.6 | 144 | 896 |

Combined Map Noise 0.61 ($\mu$K-arcmin)

[†] Polarization noise in CMB thermodynamic units

**Table 1**. Parameters of the PICO mission. The map noise is assumed white and the numbers given are for the polarized maps ($Q/U$ or $E/B$).

| Model Name (Short Name) | Dust Model | Synchrotron Model | Other Components |
|---|---|---|---|
| Planck Baseline (Baseline) | PySM d1: modified blackbody with spatially varying $T_d$ and $\beta_d$ | PySM s1: power law spectrum with spatially varying $\beta_s$ | None |
| Dust: Two Modified Black Bodies (2MBB) | PySM d4: two component dust model of [23] | PySM s3: power law spectrum with spatially varying $\beta_s$ and sky-constant curvature | PySM a2 AME model: Spatially varying spectrum with fixed 2% polarization fraction |
| Physical Dust (PhysDust) | PySM d7: physical dust model of [24] including magnetic dipole emission | PySM s3 | PySM a2 AME model |
| MHD (MHD) | Modified blackbody dust emission in each cell of a TIGRESS MHD simulation [25, 26], integrated along the line of sight | Power law synchrotron spectrum with amplitude coupled to B-fields in a TIGRESS MHD simulation [25, 26] | None |
| Multi-Layer Dust (MultiLayer) | "MKD" dust model [27] based on multiple modified blackbody emission laws in each pixel | PySM s3 | PySM a2 AME model |

**Table 2**. Summary of Polarized Foreground Models.

bands with the five models and the level of CMB B-modes as a function of $\ell$. Models 2MBB, PhysDust, and MHD, are identical to those used in [21].

### 3.1 Model Planck Baseline

The 'Baseline' model is based on polarized Galactic emission provided by the Python Sky Model [PySM2; 28], specifically the "d1" model of polarized dust emission and the "s1" model of polarized synchrotron emission. In each sky pixel the dust emission is characterized by Planck's 353 GHz polarized intensity in $Q$ and $U$ [29] and by an opacity law index $\beta_d$ and dust temperature $T_d$ to describe the frequency dependence. The $Q$ and $U$ 353 GHz amplitude maps, as well as the $T_d$ and $\beta_d$ maps, are based on the [29] component separation analysis with the Commander framework. Gaussian fluctuations are added to low-pass filtered amplitude maps at angular scales $\ell > 69$, where the data are not constraining, to produce final maps with power at scales up to $\ell \sim 1500$ [28].

The synchrotron emission in each pixel is likewise described by an amplitude in each of $Q$ and $U$ and by a power law index $\beta_s$ to describe the frequency dependence. The $Q$ and $U$ amplitudes are based on the 9-yr WMAP 23 GHz $Q$ and $U$ maps smoothed to a resolution of 3°. The $\beta_s$ map is taken from "Model 4" of [30], who used the Haslam 408 MHz survey data [31] and 3-yr WMAP 23 GHz data [32] to derive synchrotron spectral indices.

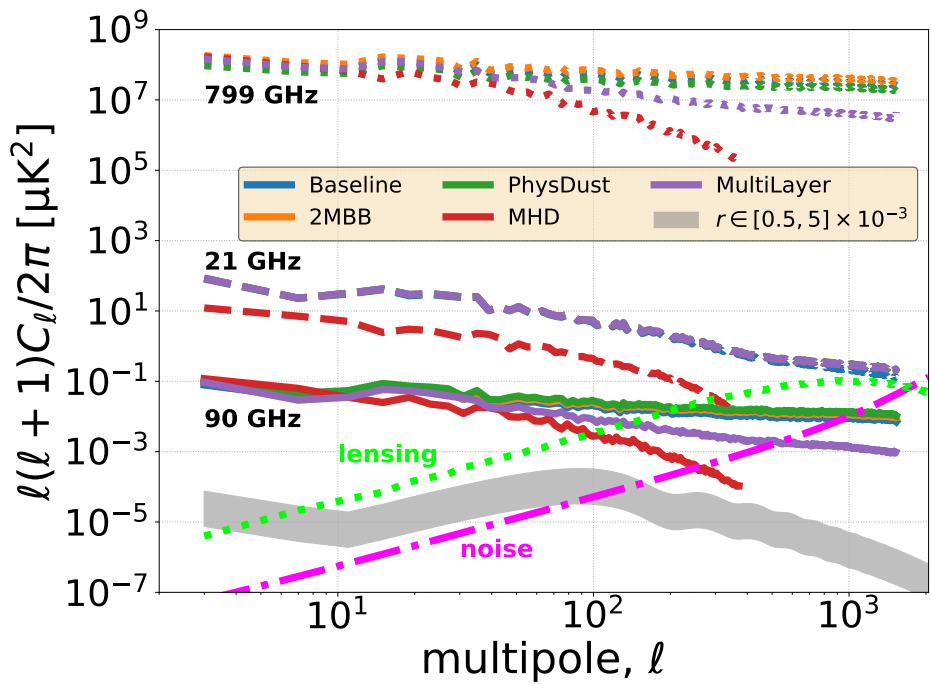

**Figure 1**. B-mode power spectra from dust and synchrotron at 21 GHz (dash), 90 GHz (solid), and 799 GHz (dots) for different foreground models on 46% of the sky compared to the primordial CMB spectrum (gray) with tensor-to-scalar ratio values $0.5 \times 10^{-3} \leq r \leq 5 \times 10^{-3}$, lensing (light green dots), and to the PICO instrument noise level (magenta dash dot). All models except MHD have information up to $\ell \simeq 1500$. Model MHD, based on a 3D MHD simulation with finite resolution, has information only up to $\ell \simeq 380$. Synchrotron emission dominates at 21 GHz and the two PySM synchrotron models used for models Baseline, 2MBB, PhysDust, and MultiLayer overlap (purple dash). Dust emission dominates at 799 GHz and the different models are constrained to match available data at low $\ell$.

## 3.2 Model Dust: Two Modified Black Bodies

Model 2MBB is also based on polarized Galactic emission models from PySM, but differs from the Baseline in the frequency scalings of both dust and synchrotron emission as well as inclusion of a polarized anomalous microwave emission (AME) component. Specifically, model 2MBB employs the "d4" dust model, "s3" synchrotron model, and "a2" AME model.

The dust emission in model d4 is based on the parametric fits of [23], who employed a two component dust model to describe a combination of Planck and DIRBE/IRAS data. The emission in each pixel is specified by an amplitude in each of $Q$ and $U$ and by a dust temperature for each of the two components to describe the frequency dependence. The temperature maps are taken directly from the fits of [23]. The $Q$ and $U$ amplitude maps are generated by scaling the model to 353 GHz in total intensity, then applying the same polarization angle and polarization fraction as used in the d1 model. Gaussian small scale fluctuations are added as in the d1 model. Thus, the 353 GHz $Q$ and $U$ amplitude maps of the d4 model differ in detail from those of the d1 model.

The s3 synchrotron model is in all respects identical to the s1 model described in the previous section with the addition of a curvature term to the frequency scaling. The synchrotron curvature parameter is taken to be constant over the full sky.

The AME emission in the a2 model is based on the parametric fits of [29], who describe the AME spectrum as the sum of two spinning dust components. The a2 model in total intensity is described by three parameter maps specifying the amplitudes of each of these components and the peak frequency of one. The peak frequency of the other is held fixed across the sky. The parameter maps are taken directly from the [29] fits. To construct $Q$ and $U$ maps, it is assumed that the AME has the same polarization angle as the [29] maps of 353 GHz polarized dust emission and a sky-constant polarization fraction of 2%.

### 3.3 Model Physical Dust

Model PhysDust is identical in all respects to 2MBB with the exception of the model for polarized dust emission, where the "d7" model is used instead of d4. Polarized dust emission in the d7 model is based on a physical model of interstellar grains with a distribution of sizes and temperatures [24]. This model notably includes grains with magnetic iron inclusions, which emit significant magnetic dipole radiation at frequencies $\lesssim 100$ GHz [33].

The $Q$ and $U$ dust amplitude maps of the d7 model are identical to those in the d1 model (employed in our model Baseline). The frequency scaling is based on a single parameter $\mathcal{U}$ quantifying the intensity of radiation heating the grains. A map of $\mathcal{U}$ is constructed from the [29] dust temperature maps assuming $\mathcal{U} \propto T_d^{4+\beta_d}$ and that $T_d = 20$ K corresponds to $\mathcal{U} \simeq 1$. The frequency scaling is based on tabulated model calculations and is not simply expressed with a parametric formula.

### 3.4 Model MHD

Unlike the Baseline, 2MBB, and PhysDust models that are taken from PySM, model MHD is based on the output of the large-scale TIGRESS magnetohydrodynamic (MHD) simulations [25]. [26] constructed full sky maps of Galactic emission from these simulations, and we use maps generated in a similar way here. Model MHD includes only polarized dust and synchrotron components.

The polarized dust emission in model MHD is based on the gas density and magnetic field orientation in each grid cell of the MHD simulation. The frequency scaling is taken to be a modified blackbody in each cell, with the dust temperature set by the local radiation field and a $\beta_d$ value having a dependence on the local gas density. The 3D simulation cube is then integrated along the line of sight to produce full sky maps following [26]. The fact that the dust frequency spectrum can vary along the line of sight introduces line of sight frequency decorrelation [27, 34, 35] in addition to decorrelation induced by spectral parameters that vary across the sky, which is present in the Baseline, 2MBB, and PhysDust models.

The polarized synchrotron emission in model MHD is based on the 3D magnetic field geometry of the simulation coupled with a simple parametric model of the cosmic ray electron number density with Galactic scale height [see 36]. The synchrotron frequency spectrum is taken to be a power law with constant spectral index across the sky.

### 3.5 Model Multi-Layer Dust

Model MultiLayer is a realization of thermal dust emission with the multi-layer dust model of [27], updated to be included in the Planck Sky Model (version 2.2.3, [37]). The key idea of the model is that if parameters describing the frequency scaling of dust emission vary across the sky, they must also vary along the line of sight. Similar to model MHD, this introduces line of sight frequency decorrelation. The total emission at 353 GHz is modelled as the sum of emissions from six dust template maps (loosely associated with six layers of distance from the observer), the sum of which is constrained to the total dust emission at this frequency in the Planck intensity and polarization maps [11, 38] as obtained with the GNILC component separation method [39].

Explicitly, in each pixel

$$Q_\nu^{\mathrm{MKD}} = \sum_k A_{d,k}^Q \left(\frac{\nu}{\nu_0}\right)^{\beta_{d,k}} B_\nu(T_{d,k}) \quad (3.1)$$

$$U_\nu^{\mathrm{MKD}} = \sum_k A_{d,k}^U \left(\frac{\nu}{\nu_0}\right)^{\beta_{d,k}} B_\nu(T_{d,k}), \quad (3.2)$$

where index $k$, running from 1 to 6, identifies a layer of emission. In practice, at high Galactic latitude only the first three layers contribute to the total emission. Closer to the Galactic plane, all six layers have non-vanishing contributions because even the more distant layer contain a significant amount of dust. These dust maps are added to PySM maps of other Galactic components to produce a full emission model enumerated dms3a2f1.

## 4  Component separation approaches and sky maps

There are two broad classes of component separation approaches, parametric and blind [e.g., 40, 41]. With the parametric approach one parametrizes the spectral dependence of the emission law for a given foreground emission component, estimates the free parameters given the data, and produces a map of the cosmological signal by marginalizing over the foreground parameters. The premise with the blind approach is not to assume any specific model for the foreground emission. Rather, one exploits the statistical independence of emission from different physical origins to separate them using observations at different frequencies. In particular, assuming that the frequency scaling of only the CMB is known, one can form linear combinations of the various observations that minimize residual foreground and noise.

We use the blind approach with all five sky models. The specific implementation is NILC [42]. We use the parametric approach only with the Baseline model, and the specific implementation is Commander1 [43, 44]. In Commander1 component separation is carried out on each map pixel. While the algorithm is computationally fast, it requires uniform angular resolution for all frequency maps. More advanced versions of Commander [45, 46] can account for beam size differences, but require significantly more compute time. For estimating $r$, most of the relevant information comes from $\ell < 150$, and for the very low values of $r$ that PICO targets and at which the lensing signal dominates, the reionization peak at $\ell \lesssim 12$ is particularly important. In this paper, the Commander analysis is restricted to $\ell \leq 12$.

We make simulated PICO-observed $Q$, and $U$ sky maps for NILC by adding the Galactic foreground models described above to realizations of CMB and experimental noise, using the same methods as used by CMB-S4 el al. [21]. Realizations of lensed-$\Lambda$CDM are borrowed from the Planck FFP10 simulations [47], as are realizations of tensor modes with $r = 0$ and $r = 0.003$. The foreground and CMB maps are smoothed assuming Gaussian beam shapes with the FWHM values given in table 1. Gaussian noise is generated in harmonic space with the levels given in table 1. The noise is assumed to have a flat spectrum in $C_\ell$. The foregrounds, CMB, and noise are added together to produce simulated observed sky maps. While there is only one realization for each of the foreground models, we add these to multiple realizations of CMB and noise. The simulations are done at a HEALPix [48] resolution $N_{\mathrm{side}} = 512$, except when we are doing lensing reconstruction and map-based delensing, for which additional maps are rendered at $N_{\text{side}} = 2048$; see section 7. The foreground component uses the same $\ell$ space information at $N_{\text{side}} = 2048$ as at $N_{\text{side}} = 512$.

Maps for Commander are constructed largely the same way as they are for NILC except for the following differences: (1) isotropic, homogeneous Gaussian noise is generated in pixel space, not harmonic space; (2) because Commander1 requires a common beam, the CMB and foreground signals have been smoothed with a Gaussian beam of 40 arcmin FWHM for all frequency bands. The noise is not smoothed; (3) as stated earlier, with Commander we only use model Baseline.

## 5 Component separation — NILC

### 5.1 Methodology

We compute the spherical harmonic transform of the PICO-observed $Q_\nu$ and $U_\nu$ full-sky maps at frequency $\nu$ to obtain the harmonic coefficients $a^E_{\ell m,\nu}$ and $a^B_{\ell m,\nu}$. For the analysis and results presented in this section we maintain only the B-mode coefficients $a^B_{\ell m,\nu}$, calculate their inverse spherical harmonic transform, and obtain full-sky B-mode maps $B_\nu$ at each frequency. The subsequent component separation process only requires the $B_\nu$ maps and its goal is to produce a best estimate of the underlying CMB B-mode map $\hat{B}^{\text{NILC}}$. However, the map-domain delensing analysis presented in section 7 also requires a best estimate component-separated CMB E-mode map $\hat{E}^{\text{NILC}}$, and to produce this map we use the $a^E_{\ell m,\nu}$ coefficients to obtain E-mode maps $E_\nu$. Although subsequent paragraphs in this section refer to component separation using $B_\nu$, the process is identical for $E_\nu$. Using full-sky maps prevents E-to-B leakage.

With NILC, the data $d_\nu$ at each frequency $\nu$ and sky pixel $p$ are assumed to be the sum of the CMB component, whose spectral energy distribution $a_\nu$ is known but fluctuation amplitude $s$ is not, and a global nuisance term $n_\nu$ which includes all other foreground emission components and instrument noise

$$d_\nu = a_\nu s + n_\nu \,. \tag{5.1}$$

In this equation and most subsequent equations the dependence on sky pixel $p$ has been suppressed to simplify notation. The data $d_\nu$ represent the observed B-mode signal at frequency $\nu$, that is $d_\nu \equiv B_\nu$. Operating on B-mode maps instead of using the Stokes $Q_\nu$ and $U_\nu$ maps optimizes foreground cleaning by having NILC minimize the variance of the foreground B-mode signal directly. The variance of the $Q$ and $U$ maps is dominated by E-modes.

The first step in the NILC algorithm is to perform a wavelet decomposition of the B-mode data on a needlet frame [49] as follows. The full-sky B-mode maps $d_\nu$ are bandpass-filtered in harmonic space using the seven window functions $(j) = (1),\ldots,(7)$ shown in figure 2. This provides seven maps $d_\nu^{(j)}$, each exhibiting fluctuations of a specific range of angular scales $(j)$, for each frequency band. For each range of scales $(j)$ we compute an estimate $\hat{s}^{(j)}$ of the CMB B-mode anisotropy $s$ in each pixel by forming a weighted linear combination of the frequency band data $d_\nu^{(j)}$

$$\hat{s}^{(j)} = \sum_\nu w_\nu^{(j)} \, d_\nu^{(j)} \,, \tag{5.2}$$

using the specific weights

$$w_\nu^{(j)} = \frac{\sum_{\nu'} \mathrm{C}^{-1\,(j)}_{\nu\nu'} a_{\nu'}}{\sum_{\nu'}\sum_{\nu''} a_{\nu'} \mathrm{C}^{-1\,(j)}_{\nu'\nu''} a_{\nu''}} \,. \tag{5.3}$$

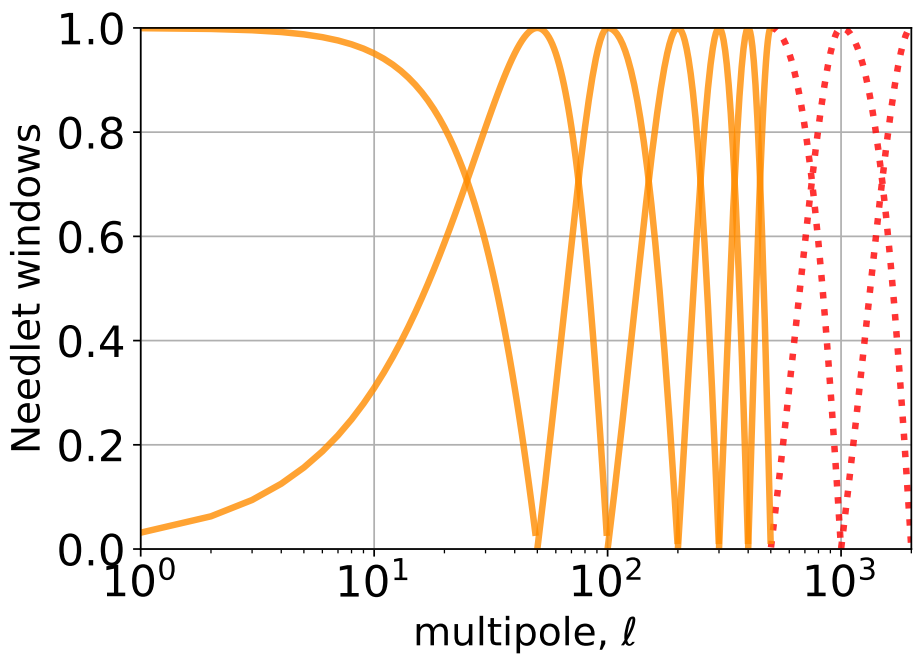

**Figure 2**. Cosine-shaped needlet bandpasses in harmonic space. Most of the analysis uses the seven bandpasses that peak at $\ell = 0, 50, 100, 200, 300, 400$, and $500$ (solid orange lines). Two additional bandpasses (red dots), one peaking at $\ell = 1000$, and another peaking at $\ell = 500$ and $\ell = 2000$ are only used for the high-resolution maps of section. 7.

$C^{-1(j)}_{\nu\nu'}$ are the elements of matrix $C^{-1(j)}$ which is the inverse of the empirical covariance matrix of the data $C^{(j)}$ in each pixel $p$ and needlet scale $(j)$, whose elements are computed as

$$C^{(j)}_{\nu\nu',p} = \sum_{p'} W^{(j)}_{p,p'} d^{(j)}_{\nu,p'} d^{(j)}_{\nu',p'}. \quad (5.4)$$

The matrix coefficient $C^{(j)}_{\nu\nu',p}$ is the result of a convolution in pixel space and for this reason we show the explicit dependence on $p$. The Gaussian convolution kernel $W^{(j)}$ in equation (5.4) defines the spatial domain over which the average of the product of maps $d^{(j)}_\nu d^{(j)}_{\nu'}$ for a given pair of frequencies is computed. The size of the convolution kernel is set by the number of independent modes within each spatial domain, which requires a FWHM of the Gaussian function $W^{(j)}$ of 4.0, 1.6, 0.85, 0.56, 0.46, 0.40, and 0.51 radians for needlet scale $(j) = (1), \ldots, (7)$, respectively. These values guarantee sufficient spatial localization of the convolution kernel in pixel space while keeping the number of pixels large enough to ensure that the so-called 'ILC bias' [42] is lower than a percent. We then combine the seven estimated CMB B-mode maps $\hat{s}^{(j)}$ from each needlet scale to form the final CMB B-mode full sky map $\hat{B}^{\mathrm{NILC}}$. The best estimate CMB E-mode map $\hat{E}^{\mathrm{NILC}}$ is produced the same way using $E_\nu$ as input. No attempt was made to optimize the needlet window functions as a function of the effectiveness of component separation; see section 8.

For each foreground model and $r$ value, we take ten realizations of CMB and noise. For each realization, $r$ value, and foreground model, we construct the map $B^{\mathrm{NILC}}$, apply the same Galactic mask corresponding to a sky fraction $f_{\mathrm{sky}} = 0.46$, and calculate the B-mode angular power spectrum $\widehat{C}^{BB}_\ell$ using MASTER [50]. The mask was built from sky model Baseline by nulling the pixels in which the variance of the 40′-smoothed observed B-mode map at 555 GHz is the largest, until 50% of the sky is masked out, see figure 3. Apodization of the mask border leaves an effective sky fraction of 46%. This mask choice has not been optimized, although we will see later that it appears to be conservative for foreground models Baseline-MHD. Power spectra error bars are derived analytically using equation (A.7). Angular power spectra of residual foreground and noise are obtained by applying the NILC weights equation (5.3) to foreground-only and noise-only maps, then following the same process that generates $\hat{B}^{\mathrm{NILC}}$.

Using the B-mode power spectrum of $\hat{B}^{\mathrm{NILC}}$ and the corresponding noise power spectrum, we calculate a likelihood for $r$ including information from multipoles $2 \leq \ell \leq 300$. We assume the Gaussian likelihood given in equation (A.6), which is a common approximation for the exact likelihood. We discuss the approximation in appendix A. To account for PICO's capability to improve $r$ constraints through delensing, we subtract 73% of the cosmic variance of the lensing B-mode signal from the covariance matrix of the likelihood, that is, in equations (A.6) and (A.7) we replace $\widehat{C}_\ell^{BB,\mathrm{NILC}}$ with $\widehat{C}_\ell^{BB,\mathrm{NILC}} - 0.73\, C_\ell^{\mathrm{lens}}$. We justify the level of 73% in section 7.1. When confidence intervals are quoted, they are calculated by integrating the likelihood to encompass 68% when $r \neq 0$, or to 95% when $r = 0$.

## 5.2 NILC results

Figure 3 gives two B-mode map-domain examples for the component separation results. The left column is for the Baseline model, which is also prototypical of the results with 2MBB, PhysDust, and MHD, and the right column is for the MultiLayer model. At 90 GHz, and at all other PICO frequency bands, the input sky map (first row) is entirely dominated by foregrounds. The input CMB map, containing no inflationary signal and only lensing induced B-modes (second row), has a scale ten times lower than the full emission map. Nevertheless, the component separation produces an output CMB map (third row) that is visually nearly indistinguishable from the input, including along the Galactic plane. With model MultiLayer, the Galactic plane is clearly visible and there is a larger variance across sky. The residual maps showing the difference of output and input (fourth row) more clearly reveal the same conclusions.

Figure 4 gives the NILC component separation results. The top row show results for each of the ten realizations with the Baseline model with the two $r_{\mathrm{in}}$ values. The recovered CMB signal matches the input, which is dominated by lensing for $\ell \gtrsim 10$; the noise post-component separation is well below the signal; and the residual foregrounds are well below the noise. The middle panels show the average power spectra over the ten realizations and demonstrate that these conclusions hold for all models except MultiLayer. The inflection point in the residual noise spectra at $\ell \sim 800$, most prominently visible for the MultiLayer model (purple triangles), is a typical consequence of NILC weighting [11]. The outcome of the algorithm is to minimize the foreground variance at low $\ell$ at the expense of higher noise. At high $\ell$ there is minimization of noise variance. For all models except MultiLayer the residual foreground spectra are equivalent to levels below $r = 5 \times 10^{-4}$ for $2 \leq \ell \leq 150$. The lower row gives likelihoods for $r$ calculated using the average spectra. With the exception of the MultiLayer model, both $r = 0.003$ and an upper limit on $r = 0$ are obtained without bias. With MultiLayer the residual foreground is larger than with other sky models and it reaches a level of $r = 0.005$ near $\ell = 10$. The large residual leads to biased posterior distributions for both $r = 0$ and $r = 0.003$. We discuss the MultiLayer model in more detail in section 5.2.1.

Table 3 summarizes the inferred values of $r$ for input values $r_{\mathrm{in}} = 0.003$ and $r_{\mathrm{in}} = 0$. For all sky models except MultiLayer, $r = 0.003$ is detected with insignificant bias and more than $5\sigma$ significance without accounting for delensing, and with $18\sigma$ significance or more with 73% delensing. Similarly, with $r_{\text{in}} = 0$ the inferred values are consistent with $r = 0$, and after delensing the 95% confidence upper bounds are between 1.3 and $2.7 \times 10^{-4}$. With model MultiLayer there is a $3\sigma$ bias with both $r$ values after 73% delensing. Residual foregrounds after component separation lead to biased estimates.

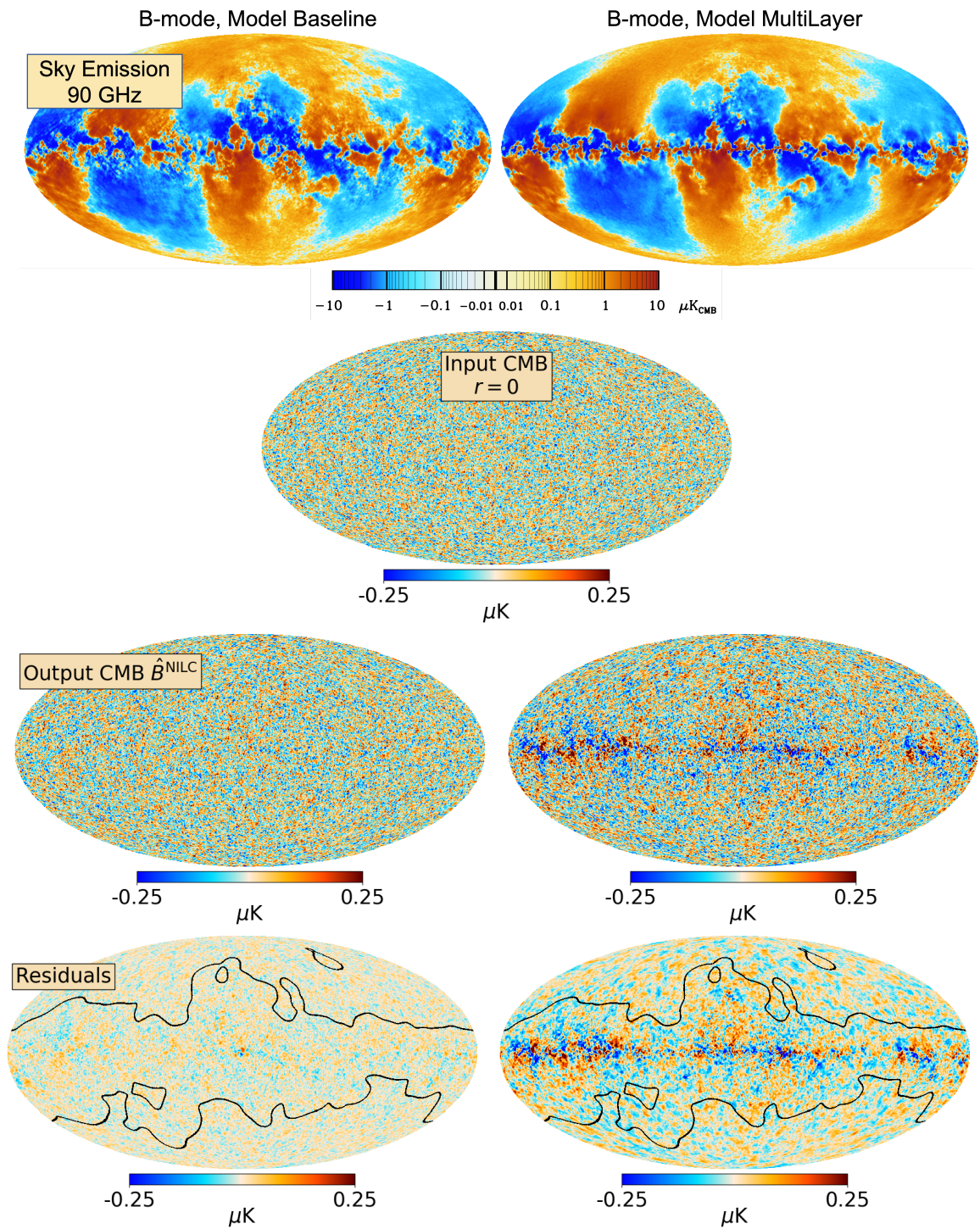

**Figure 3**. B-mode maps smoothed to 40' fwhm for models Baseline (left) and MultiLayer (right) before and after component separation with NILC. From top to bottom: sky map at 90 GHz; the input CMB with $r = 0$; the output CMB after component separation; and the CMB residual map = output CMB - input CMB, and outline of the 50% mask that determines the portion the sky for power spectra calculations.

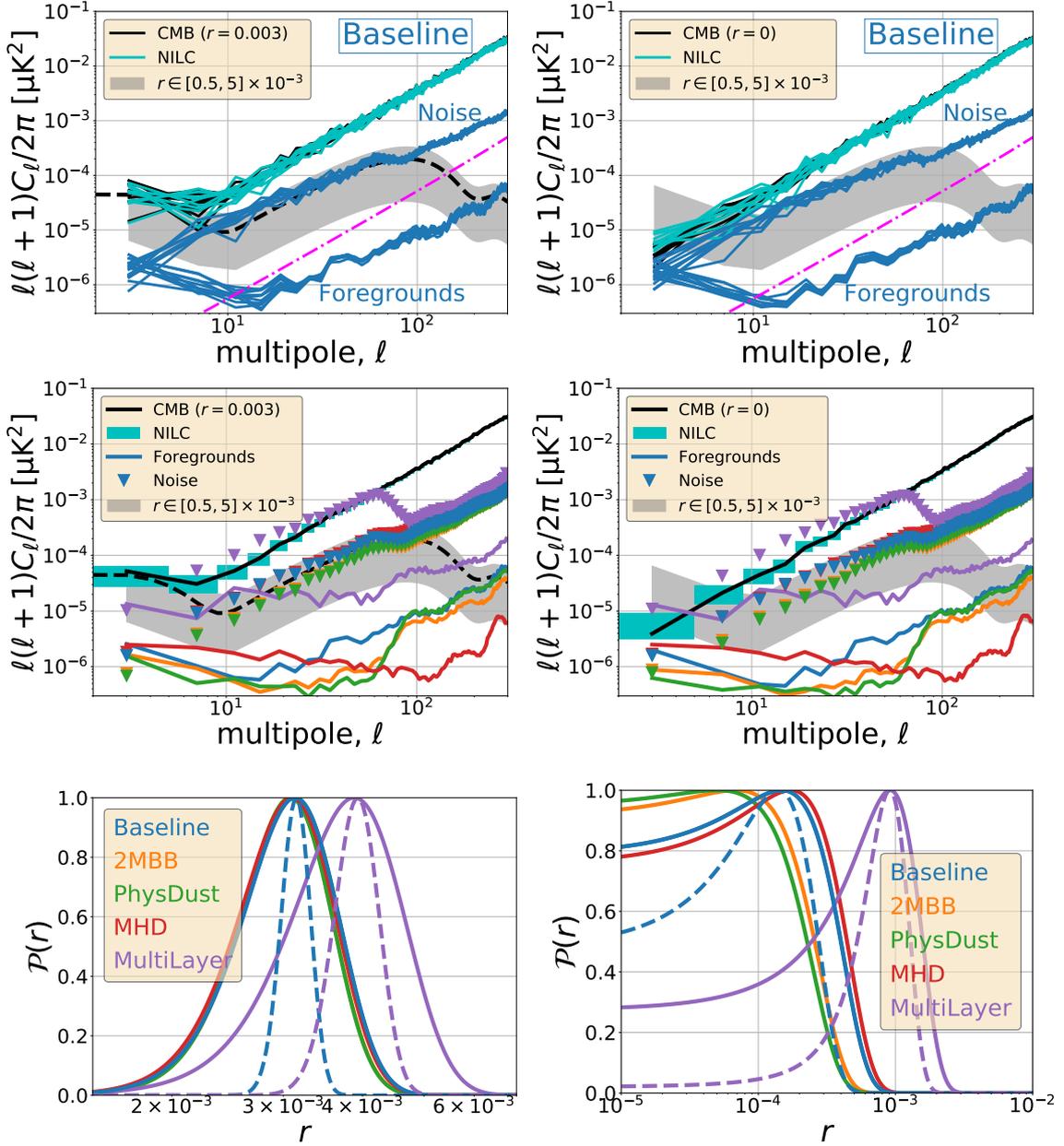

**Figure 4**. Power spectra and $r$ likelihood after NILC component separation with $r_{\rm in} = 0.003$ (left) and $r_{\rm in} = 0$ (right). Top: Recovered B-mode power spectra of CMB (cyan), residual foregrounds (blue) and noise (blue) for sky model Baseline and ten realizations of CMB and noise. For reference, the black solid lines show the power spectra of the input CMB realisations while the gray shaded area shows the theoretical power spectrum of the primordial CMB B-mode for $r$ ranging from $5 \times 10^{-4}$ to $5 \times 10^{-3}$. Middle: Power spectra averaged over all 10 realisations of residual foregrounds (solid coloured lines) and noise (coloured triangles) for all sky models: Baseline (blue), 2MBB (orange), PhysDust (green), MHD (red), and MultiLayer (purple). The cyan boxes show the average recovered CMB B-mode power spectrum and standard deviations for model Baseline. Bottom: Recovered posterior distributions of $r$ without delensing (solid) and with $A_{\rm lens} = 27\%$ for models Baseline and MultiLayer (dash).

|  | $r_{\text{in}} = 0.003$ | | $r_{\text{in}} = 0$ | | | |
|---|---|---|---|---|---|---|
|  | $[r \pm \sigma(r)]/10^{-3}$ | | $r_{95\%}/10^{-4}$ | $r/\sigma(r)$ | $r_{95\%}/10^{-4}$ | $r/\sigma(r)$ |
| Model | No delensing | 73% delensing | No delensing | | 73% delensing | |
| Baseline | $3.12 \pm 0.54$ | $3.15 \pm 0.16$ | 3.7 | $< 1$ | 2.6 | 1.2 |
| 2MBB | $3.07 \pm 0.50$ | $3.09 \pm 0.13$ | 2.5 | $< 1$ | 1.5 | $< 1$ |
| PhysDust | $3.07 \pm 0.50$ | $3.09 \pm 0.11$ | 2.3 | $< 1$ | 1.3 | 1.2 |
| MHD | $3.09 \pm 0.53$ | $3.09 \pm 0.17$ | 4.1 | $< 1$ | 2.7 | 1.2 |
| MultiLayer | $3.90 \pm 0.79$ | $3.93 \pm 0.32$ | 14.8 | 1.6 | 13.2 | 2.8 |

**Table 3.** $r$ forecasts using blind component separation with $r_{\text{in}} = 0.003$ and $r_{\text{in}} = 0$ and using 21 frequency bands. $r_{95\%}$ denotes the 95% upper limit, while $r/\sigma(r)$ gives the value of $r$ at the peak of the likelihood relative to the $1\sigma$ interval, calculated by integrating the likelihood from its peak to 68% probability.

The results presented in table 3 and in figure 4 use all the frequency bands for the component separation. We have analyzed the need for the wide frequency range by carrying out the component separation and removing some edge bands. A PICO-HF configuration assumes bands only between 43 and 799 GHz. In this configuration the four lowest PICO bands have been removed. A PICO-LF configuration assumes bands only between 21 and 462 GHz. In this configuration the three highest PICO bands have been removed; see table 1. The PICO-HF and -LF configurations are intended to give a coarse assessment about the relative importance of accounting for synchrotron vs. Galactic dust and to inform the level of constraints attainable with a narrower frequency range. A frequency range narrower than PICO's baseline are planned for next generation ground-based instruments and have been proposed for a future space mission [2, 6, 51]. Results from this analysis, which was only carried out for $r_{\text{in}} = 0$, are given in figure 5 and table 4.

For either band-restricted configuration and for all sky models the level of residual foregrounds increases (figure 5 left column), degrading constraints on $r$ and in some cases producing statistically significant biases. For example, with models Baseline and MHD, the value of $r$ at the peak of the likelihood is approximately $3\sigma$ away from $r = 0$. For sky models 2MBB and PhysDust, which include synchrotron curvature, the constraints on $r$ degrade approximately equally with PICO-HF and -LF indicating an approximate equal role for separating Galactic dust and synchrotron. In contrast, with model Baseline, which has no synchrotron curvature, or with models MHD and MultiLayer, in which Galactic dust has more complex emission properties, PICO-HF gives slightly less biased $r$ constraints compared to PICO-LF, indicating that separating Galactic dust is more important than separating the relatively simpler synchrotron emission model. Overall, a 21-band PICO instrument would reduce the 95% upper limit on $r = 0$ by 20–50% relative to PICO-HF and by 40–54% relative to PICO-LF, see table 4.

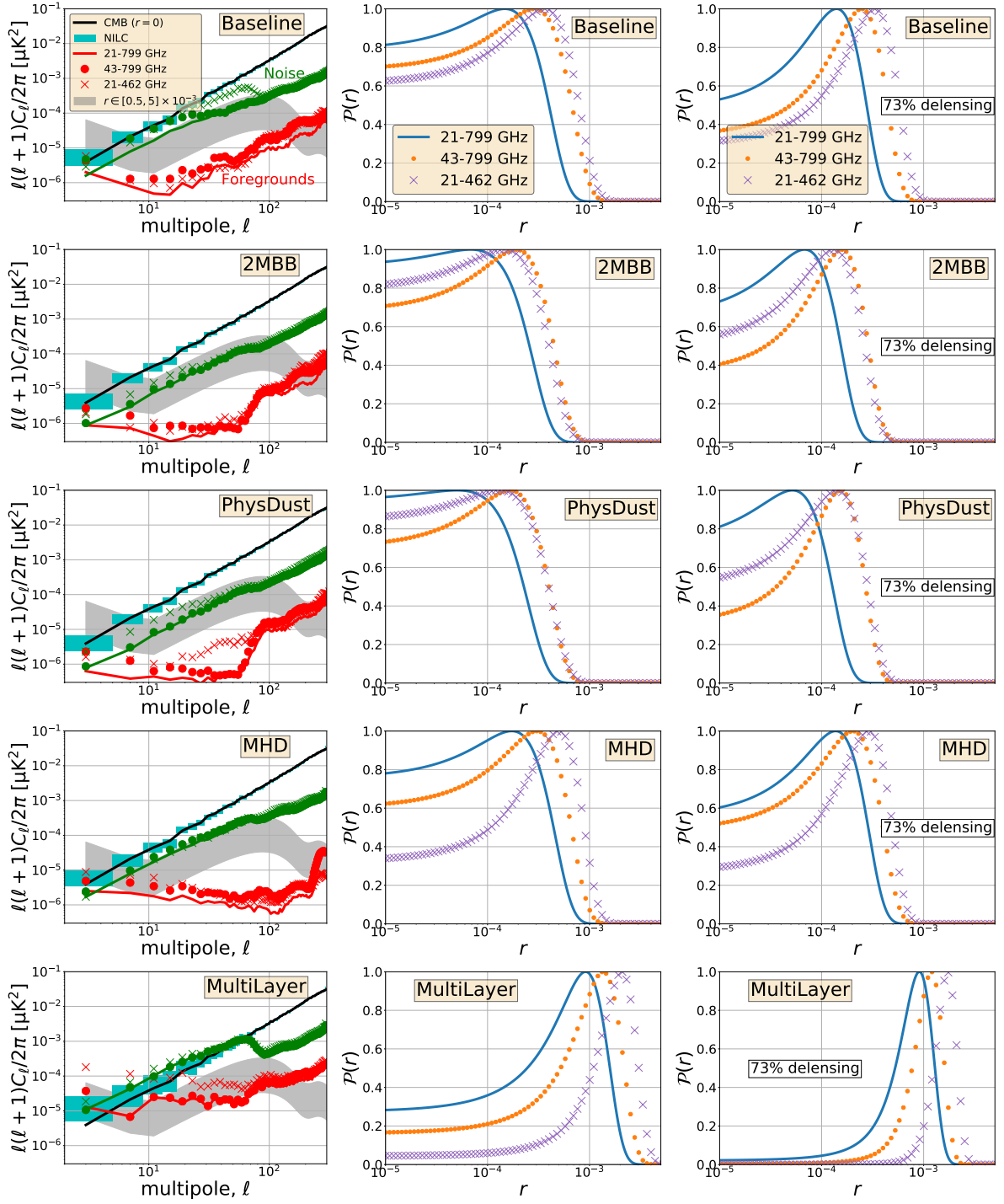

**Figure 5**. Power spectra (left column) and $r$ likelihoods without and with delensing (middle and right columns) when $r_{\rm input} = 0$ comparing three configurations: all frequency bands (21 GHz – 799 GHz, solid); without the low frequency bands (43 GHz – 799 GHz, dots); and without the high frequency bands (21 GHz – 462 GHz, crosses). Legends for the panels are given in the first row.

– 14 –

|  | $r_{\rm in} = 0$, 73% delensing | | | |
|---|---|---|---|---|
|  | PICO-HF (43-799 GHz) | | PICO-LF (21-462 GHz) | |
| Model | $r_{95\%}/10^{-4}$ | $r/\sigma(r)$ | $r_{95\%}/10^{-4}$ | $r/\sigma(r)$ |
| Baseline | 4.3 | 1.5 | 5.6 | 1.6 |
| 2MBB | 2.8 | 1.4 | 2.5 | 1.3 |
| PhysDust | 2.6 | 1.7 | 2.7 | 1.3 |
| MHD | 3.8 | 1.3 | 4.8 | 1.6 |
| MultiLayer | 16.7 | 3.4 | 22.8 | 4.4 |

**Table 4**. NILC $r$ forecasts with $r_{\rm in} = 0$ and without either the low frequencies (LF) or the high frequencies (HF), and assuming 73% delensing.

### 5.2.1 Model MultiLayer and local estimates of $r$

With model MultiLayer, the NILC component separation process gives biased estimates of $r$. In section 8 we discuss potential improvements for the component separation process that might mitigate the biases.

One way to identify the existence of biased estimates during the analysis is to compare independent constraints on $r$ from independent sections of the sky. To carry out this analysis, we partition the sky to 26 sections of equal area outside the Galactic plane, each with a 2.5% area of the sky, see figure 6. Masks for individual regions are directly applied to the $\hat{B}^{\rm NILC}$ map, with no apodization, to compute local B-mode spectra and infer local constraints on $r$. No apodization is necessary because there is no E-to-B leakage and no significant contamination of small scale power from large angular scale modes. With models Baseline and MultiLayer we plot the constraints as a function of the rms dust polarization intensity $\sigma_{P_{555}}$ as determined by the 555 GHz data in each section prior to component separation

$$\sigma_{P_{555}} = \sigma\left(P_{555} = \sqrt{Q_{555}^2 + U_{555}^2}\right), \tag{5.5}$$

where $Q_{555}$ and $U_{555}$ are the pixel values in each section. There is no difference in the component separation process itself; it is conducted on the full sky. The results are shown in figure 6 for $r_{\rm in} = 0$ (left panel) and $r_{\rm in} = 0.003$ (right panel). In both cases we show $r$ estimates after 73% delensing. For both input $r$ values we identify a common trend that is best observed by following $\sigma_{P_{555}}$ from higher to lower values. As $\sigma_{P_{555}}$ decreases the values of the estimated $r$ values on the individual sky sections are decreasing. For the lowest $\sigma_{P_{555}}$ values, those corresponding to the cleanest sections of the sky, $r$ converges to a stable value; That is, the same value or upper limit is obtained within statistical errors. The trend indicates that, as expected, sky areas with high polarized intensity from dust correlate with biased $r$ values. As the dust polarization intensity decreases, the bias decreases and at a certain level becomes insignificant.

When $r_{\rm in} = 0$, most patches give upper limits at 95% confidence (bars with arrows) even when dust contamination is more pronounced. However, with model MultiLayer, which has higher dust residual after component separation than Baseline, four patches give biased $r \neq 0$ detections at levels exceeding $2\sigma$ (red thin error bars). When $r_{\rm in} = 0.003$ the higher dust residual of model MultiLayer gives higher $r$ levels at high $\sigma_{P_{555}}$ and larger error bars compared to Baseline. The final result is that in both models there are five patches that give

– 15 –

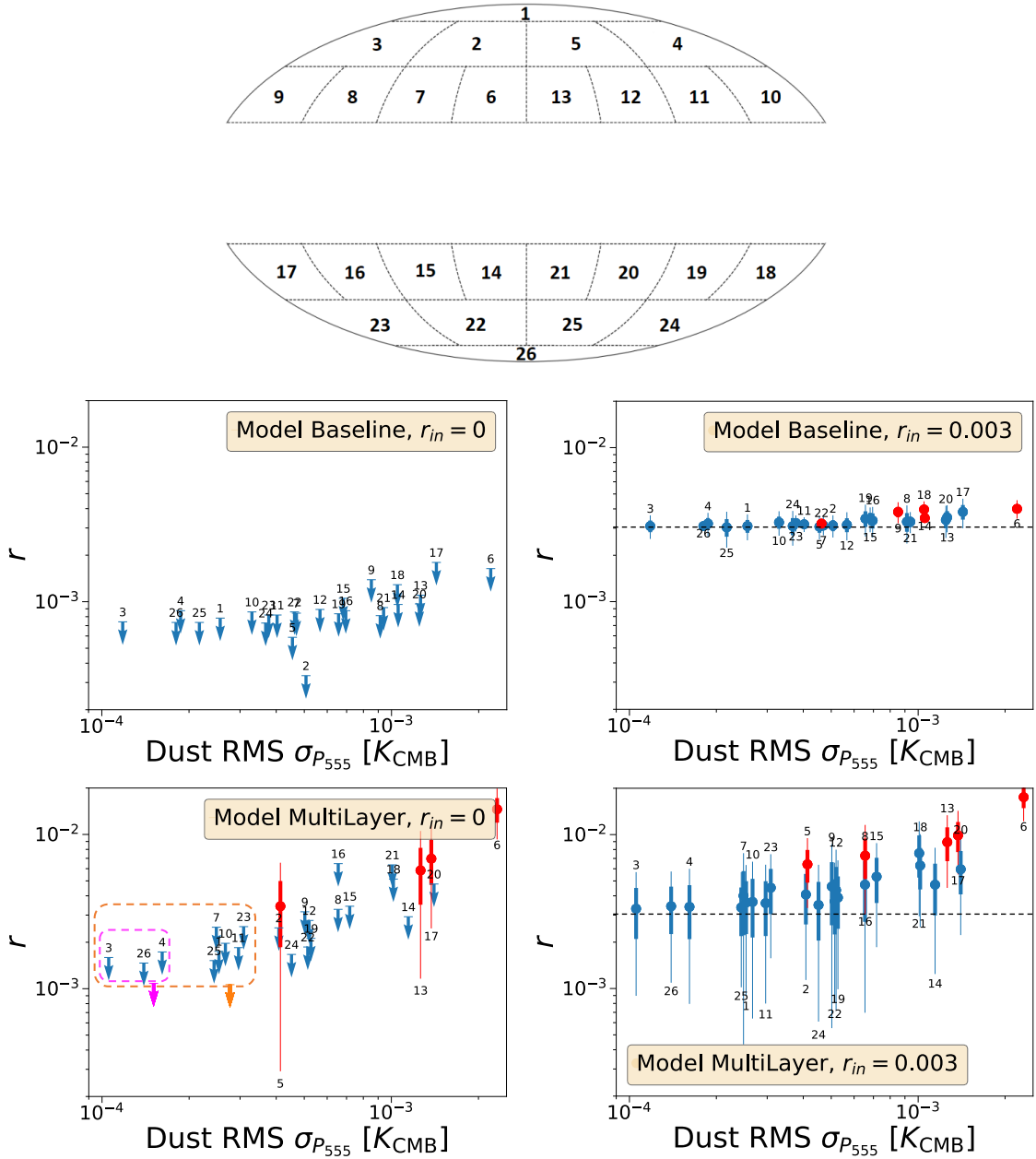

**Figure 6**. The top panel shows the 26 equal area sky sections with $f_{\rm sky}= 2.5\%$ each. The middle and bottom rows show the section-by-section results from NILC for models Baseline and MultiLayer, respectively. The left and right panels give 95% confidence limits (blue bars with arrows) with $r_{\rm in} = 0$, and $r$ likelihood confidence intervals (thick blue is 68%, thin blue is 95%) with $r_{\rm in} = 0.003$ (black dashed line), respectively, both as a function of $P_{555}$. Some patches produce biased detections at a confidence level larger than 95% (red points), both when $r = 0$ (lower left, for Multilayer) and when $r = 0.003$ (right column, Baseline and Multilayer). For biased detections we give the 68% (thick red) and 95% (thin red) confidence intervals. With model MultiLayer and $r_{\rm in} = 0$ we also give 95% upper limits after aggregating the sections with the three lowest values (lower left, magenta arrow) and with the nine lowest values (lower left, orange arrow) of dust polarization. The vertical scale in all panels is identical to give a visual comparison of the results.

biased detections of $r > 0.003$ by more than $2\sigma$ significance, and in both models many lower dust level patches are consistent with $r_{\rm in} = 0.003$. Most of the patches that give erroneous detections are near the Galactic plane, although not all, and all are at levels of 4 times the lowest $\sigma_{P_{555}}$ or above.

Aggregating several areas of the sky that have low dust polarized intensity, and for which $r$ upper limits have converged, decreases sample variance and decreases the upper limit. By aggregating the three (magenta) and nine lowest $P_{555}$ (orange) sky areas we find with model MultiLayer 95% upper limits of $1.9 \times 10^{-3}$ and $1.6 \times 10^{-3}$, respectively.

## 6 Component separation — Commander

### 6.1 Methodology

The Commander data model $d_\nu$ at each frequency $\nu$ and sky pixel is given by the sum of the signal components and a noise term $n_\nu$,

$$\begin{aligned}
d_\nu &= a_{\rm CMB} \\
&+ a_{\rm s}\, \gamma(\nu) \left(\frac{\nu}{\nu_{0,\rm s}}\right)^{\beta_{\rm s}} \\
&+ a_{\rm d}\, \gamma(\nu) \frac{e^{\frac{h\nu_{0,d}}{kT_{\rm d}}} - 1}{e^{\frac{h\nu}{kT_{\rm d}}} - 1} \left(\frac{\nu}{\nu_{0,d}}\right)^{\beta_{\rm d}+1} \\
&+ n_\nu.
\end{aligned} \quad (6.1)$$

At each pixel the free parameters are the CMB, synchrotron, and thermal dust amplitudes, $\{a_{\rm CMB}, a_{\rm s}, a_{\rm d}\}$; the synchrotron and thermal dust spectral indices, $\beta_{\rm s}$ and $\beta_{\rm d}$; and the dust temperature, $T_{\rm d}$. In addition, $k$ and $h$ are Boltzmann's and Planck's constants, respectively, $\gamma(\nu) = (e^x - 1)^2/(x^2 e^x)$ with $x = h\nu/kT_{\rm CMB}$, is the unit conversion factor between Rayleigh-Jeans brightness temperature and the CMB thermodynamic temperature, and $\nu_{0,\rm s}$ and $\nu_{0,\rm d}$ are reference frequencies for the synchrotron and thermal dust components.

The aim is to fit all the modelled parameters $\omega = \{a_i, \beta_i, T_d\}$ to the input data, which in a Bayesian framework means computing the posterior distribution $P(\omega|d)$, where $d = \{d_\nu\}$. With Bayes' rule this is given as the product of the likelihood $P(d|\omega)$ and a set of priors $P(\omega)$. The amplitudes $a = \{a_{\rm CMB}, a_{\rm d}, a_{\rm s}\}$ and spectral indices $\theta = \{\beta_{\rm s}, \beta_{\rm d}, T_{\rm d}\}$ are fitted with Gibbs sampling iteratively in steps described by

$$\begin{aligned}
a^{i+1} &\leftarrow P(a|\theta^i, d) \\
\theta^{i+1} &\leftarrow P(\theta|a^{i+1}, d)
\end{aligned} \quad (6.2)$$

where the parameters to the left of $\leftarrow$ are sampled from the distribution to the right, keeping the spectral indices fixed while sampling the amplitudes and vice versa. As discussed by [44], the amplitude distribution $P(a|C_l^i, \theta^i, d)$ is a simple multi-variate Gaussian in the various component amplitudes, while the spectral index distribution $P(\theta|C_l^i, a^{i+1}, d)$ can be sampled efficiently by mapping out the corresponding one-dimensional distribution by brute force per pixel and parameter. We adopt a standard Blackwell-Rao estimator for likelihood and cosmological parameter estimation [52].

Commander is first run for the full sky, fitting component amplitudes and spectral parameters at each pixel. We define a per-pixel $\chi^2$ statistic

$$\chi^2 = \sum_\nu \left(\frac{d_\nu - s_\nu}{\sigma}\right)^2, \qquad (6.3)$$

where $s_\nu$ is the total signal model and $\sigma$ are the instrumental noise rms per pixel. This $\chi^2$ map is smoothed to $8°$ FWHM and conservatively thresholded at $4\sigma$ to remove pixels with a significant model errors. The resulting confidence mask, outside of which the CMB is presumed reconstructed at higher confidence, removes 25 % of the low-latitude and high-foreground sky; it is shown in gray in figure 7.

We use a Blackwell-Rao estimator [52] to estimate $r$. The estimator requires a series of full-sky CMB map samples, rather than the partial sky samples as are available at this stage. We therefore run another Gibbs chain for just the CMB map of the form,

$$\begin{aligned} a_{\text{CMB}}^{i+1} &\leftarrow P(a_{\text{CMB}}|d, a_{\text{s}}, a_{\text{d}}, \theta, C_l^i) \\ C_l^{i+1} &\leftarrow P(C_l^i|a_{\text{CMB}}^{i+1}) \end{aligned} \qquad (6.4)$$

in which the presence of the power spectrum $C_l$ in the first distribution ensures that the masked pixels in the CMB sky map sample are effectively "in-painted" with a constrained Gaussian realization [43]. The second distribution $P(C_l|a_{\text{CMB}}^{i+1})$ is an inverse Wishart distribution, which has a known, simple sampling algorithm [53]. The explicit expression for the Blackwell-Rao estimator given these full-sky CMB samples is given in appendix A. No correction for delensing is done with the Commander analysis, and to the extent that comparison is done with NILC, the comparison should be with the no-delensing results.

In principle, the two Gibbs chains described by equations (6.2) and (6.4) can be merged into one chain with three steps. However, splitting it into two independent steps has several advantages [54, 55]. The most important concerns the amplitude sampling step in equation (6.2). Solving for both foreground amplitudes and a power spectrum-constrained CMB component leads to a computationally costly non-local linear system with a high condition number, and increases the total runtime of the full algorithm by orders of magnitude [45]. A second advantage is that the $\chi^2$-based CMB confidence mask can be defined after the component separation step, but before estimating the CMB power spectrum. The main disadvantage of a split chain is a slightly higher white noise level, as the foreground amplitudes are allowed to explore a larger posterior volume in the first step, when the CMB component is unconstrained by the power spectrum [54]. A total of ten simulations with independent CMB and noise realizations are processed with this algorithm for both $r = 0$ and $r = 0.003$.

### 6.2 Commander results

The upper panels in figure 7 show the Stokes $Q$ and $U$ for one single reconstructed CMB realization with model Baseline, smoothed to an effective resolution of $5°$ FWHM. Commander was only used with model Baseline. The gray region is the confidence mask. The bottom panel, which shows the difference between the reconstructed and the true input CMB map with a color scale that is hundred times smaller than the signal, demonstrates the high signal-to-noise ratio maps that PICO will generate. At a visual level the residual map appears statistically consistent with white noise, and there are no discernible foreground residuals around the mask edge.

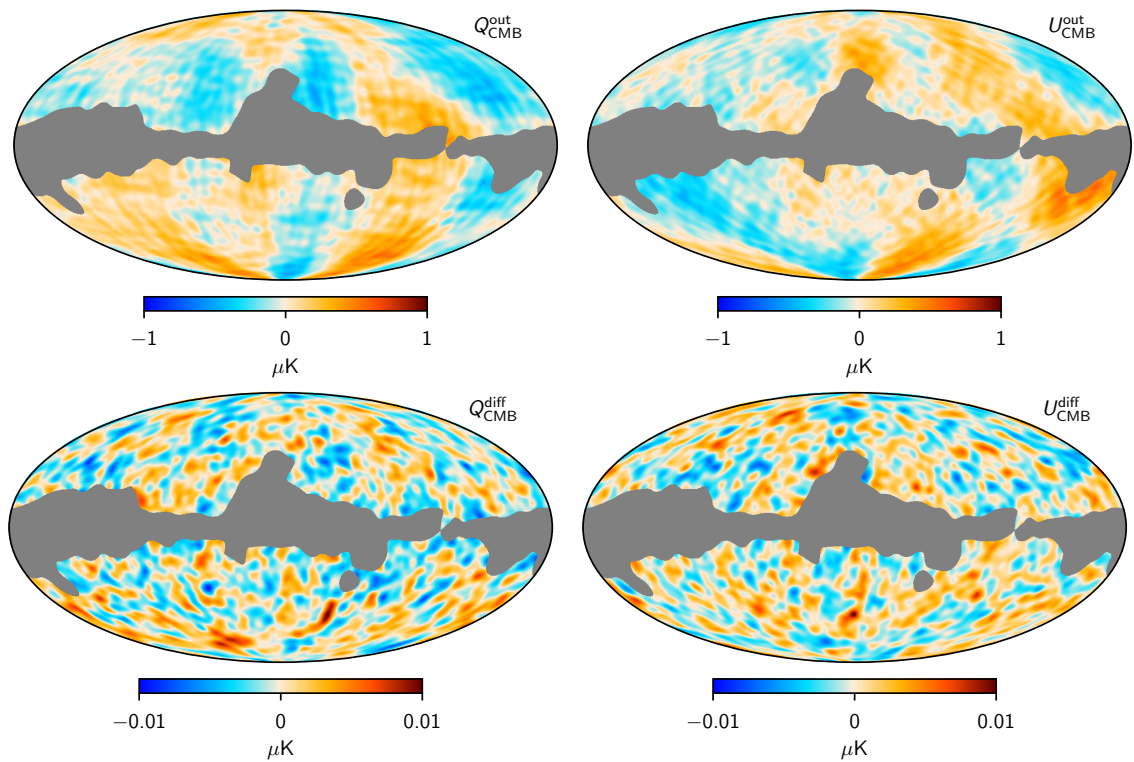

**Figure 7**. Top row: Stokes $Q$ (left) and $U$ (right) maps reconstructed with Commander for one realization with model Baseline. Bottom row: Difference between the reconstructed map and the input CMB map. The gray region is the confidence mask derived from the $\chi^2$ cut. All maps are smoothed to a common angular resolution of 5° FWHM. Note that the color scale in the bottom row is 100 times smaller than the top row.

The likelihood results from Commander are summarized in figure 8. The left and right columns shows $r$ constraints for simulations with input values of $r = 0$ and $r = 0.003$, respectively. The top row gives individual marginal posterior distributions for ten realizations of each of the $r$ inputs. In both cases, the recovered distributions are consistent with the true inputs; for $r = 0$, three out of ten realizations peak at $r = 0$, while the remaining seven realizations peak at a slightly positive value, but none has a likelihood ratio greater than $\mathcal{L}(r_{\max})/\mathcal{L}(r=0) \approx 3$, as expected for random variations. The probability of observing seven realizations with a positive value by random chance is 17 % from pure binomial statistics. Similar conclusions are obtained for the $r = 0.003$ case, with four (six) realizations having a lower (higher) peak value than the true input.

The bottom left panel of figure 8 shows the cumulative distribution functions for each of the ten realizations. The black vertical dashed line shows the median upper limit of $r < 8.9 \times 10^{-4}$ (95 % confidence limit). This limit accounts for the non-Gaussian shape of the posterior distributions, and it is a factor of 2.4 higher than the corresponding NILC limit, which uses a Gaussian likelihood. The NILC estimate, however, includes a factor of $\sim 20$ broader multipole range, $\ell < 200$, and is therefore expected to give tighter constraints. The bottom right panel shows the posterior standard deviation $\sigma_r$ when $r_{\rm in} = 0.003$ for each of the ten realizations, and the dashed black line shows their mean of $\sigma_r = 1.1 \times 10^{-3}$. For comparison, the dashed red line shows the corresponding NILC estimate for model Baseline without delensing $\sigma = 0.54 \times 10^{-3}$.

– 19 –

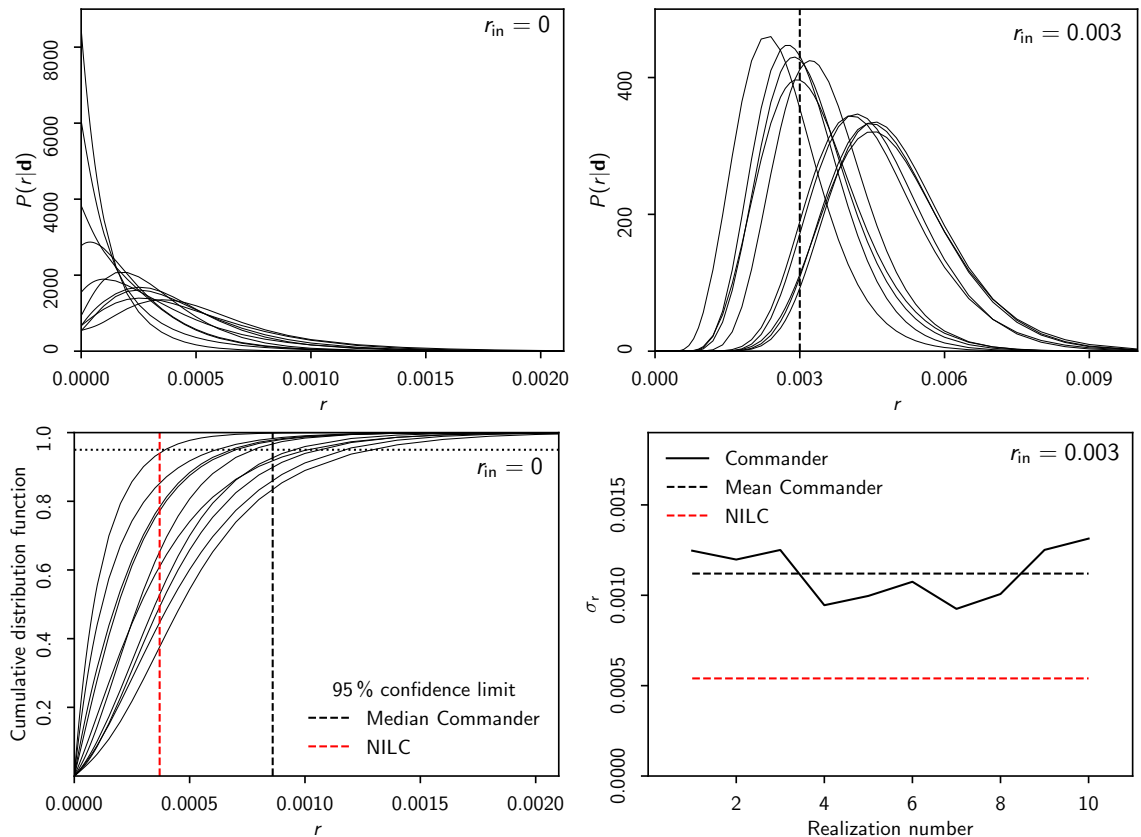

**Figure 8**. Top row: Commander posterior distributions for $r_{\rm in} = 0$ (left) and $r_{\rm in} = 0.003$ (right) with ten realizations. Bottom left: Cumulative probability distributions for ten realizations with $r = 0$. The vertical dashed black line shows the median of the 95 % confidence limits derived with Commander, while the vertical red line shows the corresponding 95 % confidence limit computed with NILC and a Gaussian likelihood. Bottom right: Posterior standard deviation $\sigma_r$ for each of the ten Commander realizations with $r = 0.003$ (solid black). For comparison, the dashed black line shows the corresponding mean, while the dashed red line shows the corresponding estimate derived with NILC coupled to a Gaussian likelihood (red dash). Note that NILC results use all simulations to generate one combined estimate while the Commander results are based on an independent estimate per realization.

## 7 Lensing and delensing

PICO constraints on $r$ depend on the effectiveness of delensing. In section 5 we assume $A_{\rm lens} = 0.27$, amounting to a delensing factor of 73%. This estimate is based on an angular power spectrum domain calculation rather than on an actual map domain delensing exercise. In section 7.1 we explain how $A_{\rm lens}$ was estimated. For one foreground model we have also conducted map domain delensing. We describe it in section 7.2 and show that the map domain delensing level is in close agreement with the power spectrum-based estimate.

### 7.1 Power spectrum domain estimate of delensing

Given PICO's noise level and beam size we forecast the beam-deconvolved noise power spectrum of the CMB polarization after component separation. For this, we use a harmonic-domain internal linear combination (ILC) approach identical to the one described by [20], apart from minor changes described in the following. The sky model includes analytic models for the polarized power spectra of dust, synchrotron, the dust-synchrotron cross-correlation, radio sources, and CMB [20]. The amplitudes of the dust and synchrotron power spectra at $\ell = 80$ are chosen to approximately match those measured on large sky fractions by Planck [56] at 353 and 30 GHz, respectively. The assumed correlation coefficient of the dust and synchrotron fields is 40%. The computed E- and B-mode power spectra at all PICO frequencies are then passed through the harmonic ILC code to obtain the post-component-separation noise on the CMB component. On the largest angular scales $\ell < 30$, we use inverse-covariance-weighted noise curves rather than the post-ILC noise, due to potential ILC biases at these low multipoles [42]; this has negligible impact on the derived lensing noise curves, and thus on the delensing performance. We implement standard minimum-variance ILC as well as constrained ILC methods [57] that deproject components with a fiducial polarized dust SED, assumed to be a modified blackbody with $\beta_{\rm d} = 1.59$, a fiducial synchrotron SED that is a power-law in antenna temperature with $\beta_{\rm s} = -3.1$, or both. This deprojection leads to larger post-ILC noise, but would significantly reduce biases due to foregrounds, and thus conservatively brackets the expected post-ILC sensitivity.

| ILC Assumptions | Map Noise Level ($\mu$K arcmin) | |
|---|---|---|
| | 0.87 | 0.61 |
| No foregrounds | 0.80 | 0.85 |
| No deprojection; standard ILC | 0.80 | 0.84 |
| Polarized dust deprojected | 0.80 | 0.84 |
| Polarized synchrotron deprojected | 0.73 | 0.78 |
| Polarized dust & synchrotron deprojected | 0.73 | 0.78 |

**Table 5**. Forecast delensing factor $1-A_{\rm lens}$ with two map noise levels for different ILC analysis assumptions [20, 57]. The delensing factor is defined in equation (A.5).

Given the post-ILC polarization noise power spectra, we use the forecasting approach of [58] to obtain the expected delensed level of the lensing B-mode power spectrum. This method involves iteratively computing the quadratic estimator-based lensing noise and then the B-mode power after delensing with the associated lensing map. With this entirely angular power spectrum-based forecasting approach, we necessarily neglect the spatial variation of the foregrounds, both for the initial harmonic-ILC step and for the subsequent iterated steps of forecasted lensing reconstruction and delensing.

Table 5 shows the delensing factor $1-A_{\rm lens}$, defined in equation (A.5), as a function of deprojection choices for two values of map noise. The lower noise is the value we use in this paper; see table 1. The higher noise is the level PICO is required to achieve [4]. We emphasize that the harmonic ILC procedure uses each band's noise level and beam size, and that the combined map noise levels are only a short-hand to indicate either configuration. The NILC component separation of section 5 conservatively uses $1-A_{\rm lens}= 0.73$, obtained when deprojecting both synchrotron and dust components with a noise level that is $\sqrt{2}$ higher than assumed in this paper. With the level of noise we use here we expect $1-A_{\rm lens}= 0.78$. In the next section, we describe a map-based delensing exercise that give $A_{\rm lens}$ values closely reproducing the results of the power spectrum-based forecasting approach.

## 7.2 Map domain delensing

To perform map-domain delensing we made maps of model Baseline with $r = 0$ at $N_{\text{side}}=$ 2048 (see section 4). We conducted NILC component separation as described in section 5, this time with an output resolution of $8'$, and the high-resolution output E- and B-mode CMB maps $\hat{E}^{\text{NILC}}$ and $\hat{B}^{\text{NILC}}$ were used to reconstruct the lensing field $\hat{\phi}$ and to produce a best estimate delensed E-mode map. This map and the field $\hat{\phi}$ are then used to estimate two additional maps, the delensed, primordial CMB B-mode map and a B-mode lensing map. Ten simulations have been conducted beginning with generation of $\hat{E}^{\text{NILC}}$ and $\hat{B}^{\text{NILC}}$ high resolution maps and culminating with these four products, one lensing field, delensed E- and B-mode maps, and a lensing map.

The lensing field $\hat{\phi}$ and delensed E-mode maps were reconstructed in two ways: (1) using a standard quadratic estimator, the output of which we denote $\hat{\phi}^{\text{QE}}$ and $\hat{E}^{\text{NILC,QE}}$ [16, 17, 59]; and (2) using a maximum a-posteriori reconstruction, the output of which is denoted $\hat{\phi}^{\text{MAP}}$ and $\hat{E}^{\text{NILC,MAP}}$ [15]. For the quadratic estimator we use Plancklens[1] and the same methodology implemented with Planck [60, 61], and the maximum a-posteriori reconstruction is obtained from a curved-sky implementation of the iterative delensing solver described by [62]. In the next paragraph we discuss only key elements of this solver and refer the reader to the cited publication for more details.

### 7.2.1 Reconstructing the lensing field $\hat{\phi}^{\text{MAP}}$ and the delensed E-mode map $\hat{E}^{\text{NILC,MAP}}$

The posterior $p$ for the lensing potential $\phi$ is

$$\log\left[p(\phi|\hat{E}^{\text{NILC}},\hat{B}^{\text{NILC}})\right] = \log\left[p(\hat{E}^{\text{NILC}},\hat{B}^{\text{NILC}}|\phi)\right] - \frac{1}{2}\sum_{LM}\frac{|\phi_{LM}|^2}{C_L^{\phi\phi,\text{fid}}}, \tag{7.1}$$

where $\phi$ is assumed to be a Gaussian field, the fiducial lensing field power spectrum $C_L^{\phi\phi,\text{fid}}$ comes from the Planck FFP10 simulations [47], and $\phi_{LM} =$ is the harmonic transform of $\phi$. The prior is necessary to handle the large number of poorly-constrained, high-$L$ modes, and it is desirable so as to optimally weigh each lensing multipole when building the lensing map [63]. We assume that the lensing likelihood $p(\hat{E}^{\text{NILC}},\hat{B}^{\text{NILC}}|\phi)$ is Gaussian in the maps. It contains a noise covariance matrix on which we apply a Gaussian isotropic beam with 8 arcmin FWHM. The noise is modeled empirically by using smoothed power spectra of $\hat{E}^{\text{NILC}}$ and $\hat{B}^{\text{NILC}}$ noise maps.

All $\hat{E}^{\text{NILC}}$ multipoles $2 \leq \ell \leq 2000$ and $\hat{B}^{\text{NILC}}$ multipoles $200 \leq \ell \leq 2000$ are used for the reconstruction of $\hat{\phi}^{\text{MAP}}$ and the delensed E-mode maps. The $\hat{B}^{\text{NILC}}$ multipole range $\ell \leq 200$ is excluded to minimize statistical dependence with the degree-scale information used for $r$-inference [64, 65]. This exclusion produces a 2% increase in the final residual lensing amplitude, that is, a 2% increase in $A_{\text{lens}}$ [66]. We assume that the unlensed CMB and the lensing map are limited to $\ell \leq 2500$ and $L \leq 2000$, respectively, and the reconstruction is done on the full-sky, since the $\hat{E}^{\text{NILC}}$ and $\hat{B}^{\text{NILC}}$ maps show no obvious spatial features which would have to be masked, see figure 9.

The solver is iterative. It calculates the gradient of the log-posterior with respect to $\phi$ and combines this information with an estimate of the local posterior curvature, which is built from previous solutions, to produce the next $\phi$ estimate [62]. Each calculation of the maximum a-posteriori reconstruction is equivalent to obtaining the most probable, or Wiener-filtered (WF), unlensed CMB conditioned on the current estimate of the lensing map and the fiducial cosmology being the truth [62].

---
[1]https://github.com/carronj/plancklens.

The likelihood model assumes $r = 0$, and we ignore temperature anisotropy information. The iteration of the solver starts from the lensing potential estimated by the QE estimator $\hat{\phi}^{0,\text{MAP}} = \hat{\phi}^{\text{QE}}$. Two maps are produced at each step of the iteration $\hat{E}^{i,\text{WF}}$ and $\hat{\phi}^{i,\text{MAP}}$, and the iteration continues ten times, a number that was verified with a separate set of simulations to produce solver convergence. Construction of $\hat{E}^{i,\text{WF}}$ is the most computationally demanding task. We use the same multigrid-preconditioned conjugate-gradient solver from the Planck analysis [61, 67], which we modified to account for the maximum a-posteriori iterative solver. The last iteration produces the best estimate $\hat{\phi}^{\text{MAP}}$, and the final Wiener filtered E-mode map is the best estimate for the delensed E-mode, namely $\hat{E}^{\text{last,WF}} = \hat{E}^{\text{NILC,MAP}}$.

### 7.2.2 Estimating the primordial and lensing B-mode maps

We produce two versions of each pair of the following: (1) a predicted B-mode lensing map, and (2) a delensed CMB B-mode map. One pair is based on $\hat{\phi}^{\text{QE}}$ and $\hat{E}^{\text{NILC,QE}}$ coming from the quadratic estimator, and a second pair is based on $\hat{\phi}^{\text{MAP}}$ and $\hat{E}^{\text{NILC,MAP}}$, which are the result of the maximum a-posteriori solver. We produce both pairs to compare their delensing performance. The B-mode lensing map is estimated by calculating the effect of the lensing field $\hat{\phi}$ on the estimated primordial, delensed E-mode field. The delensed (DL) CMB B-mode map is that obtained by subtracting the lensing B-mode map from $\hat{B}^{\text{NILC}}$, that is,

$$\hat{B}^{\text{QE,DL}} = \hat{B}^{\text{NILC}} - \hat{B}^{\text{QE,L}}, \qquad \hat{B}^{\text{MAP,DL}} = \hat{B}^{\text{NILC}} - \hat{B}^{\text{MAP,L}}, \qquad (7.2)$$

where $B^{\text{QE,L}}$ and $B^{\text{MAP,L}}$ are the lensing maps, which some authors refer to as 'lensing templates' [68, 69]. When $r \neq 0$ and for an ideal full sky, high sensitivity, and high angular resolution experiment the DL maps would approach the true primordial signal. When $r = 0$ they would only include noise. In practice, delensing is never complete and the DL maps have residual lensing. Because we have access to the simulated input lensing map we assess the residual lensing (RL) for each of the lensing estimators. We make two maps of residuals

$$\hat{B}^{\text{QE,RL}} = B^{\text{L}} - \hat{B}^{\text{QE,L}}, \qquad \hat{B}^{\text{MAP,RL}} = B^{\text{L}} - \hat{B}^{\text{MAP,L}}, \qquad (7.3)$$

where $B^{\text{L}}$ are the simulated input lensing maps. With the RL maps we calculate the residual lensing amplitude $A_{\text{lens}}$. For each delensing estimator we calculate power spectra post-delensing for full sky maps and for maps with the NILC mask. After deconvolution from the mask (following [70]), the spectra of the ten simulations are averaged to produce an average spectrum.

### 7.2.3 Map delensing results

Figure 9 shows the pre- and post-delensed B-mode maps as produced by the iterative maximum a-posteriori solver. They are compared to the input B-mode lensing $B^{\text{L}}$ and the map of residual lensing $\hat{B}^{\text{MAP,RL}}$. The $\hat{B}^{\text{NILC}}$ map pre-delensing (top left) is a high resolution version of the map shown for model Baseline in figure 3 (left column, third panel). The map variance is $\sigma^2 = (0.086\,\mu\text{K})^2$, where $\sigma$ is the full-sky pixel standard deviation of the maps band-passed to $2 \leq \ell \leq 200$, with the dominant contribution $\sigma^2 = (0.082\,\mu\text{K})^2$ coming from the input lensing map, shown for a small patch of sky at the bottom left panel; 9% of the total variance comes from residual foregrounds and noise. The post-delensing variance decreases to $\sigma^2 = (0.047\ \mu\text{K})^2$ (top right), of which $(0.039\ \mu\text{K})^2$, or 69%, is due to the residual lensing and the noise introduced in the delensing process.

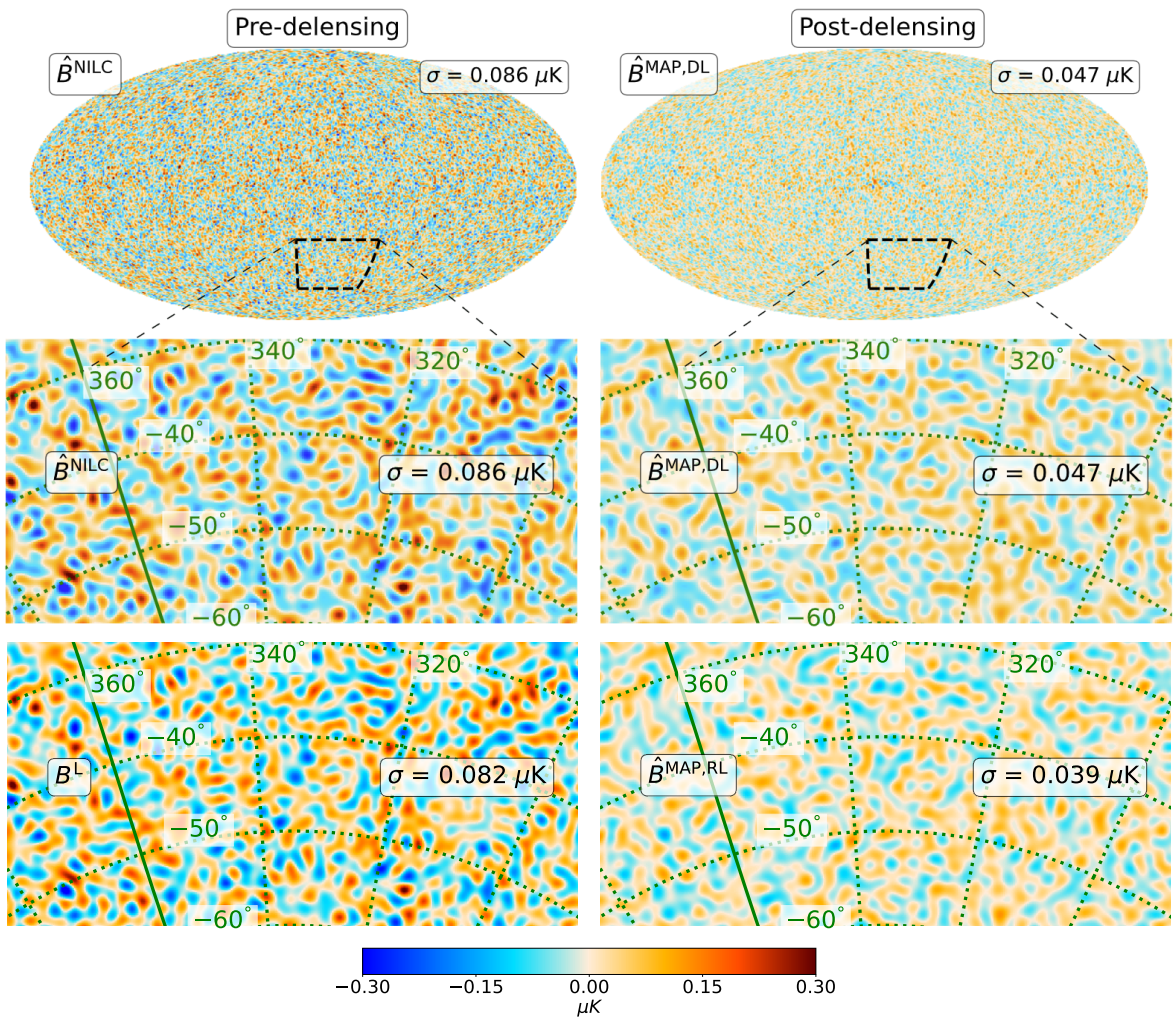

**Figure 9**. B-mode maps in Galactic coordinates after NILC component separation for model Baseline and $r_{\rm in} = 0$, before (left) and after (right) iterative delensing including full sky (top panels), smaller ($60° \times 30°$) area centered on a clean patch of the sky (middle panels), the input B-lensing map $B^{\rm L}$ (bottom left), and residual lensing map (bottom right) in the same area. All maps are band-passed to only show $2 \leq \ell \leq 200$, and no masking has been applied to the lensing reconstruction. The quantity $\sigma$ in all maps is the full-sky pixel standard deviation of the band-passed maps.

The reduction of map variance by a factor of 3 to 4 post-delensing is also apparent when comparing the angular power spectra of the pre-delensing map $\hat{B}^{\rm NILC}$ and the post-delensing map $\hat{B}^{\rm MAP,DL}$, see figure 10. The results shown in the figure are equivalent to the first row of figure 5 except that here the maps are rendered at high resolution (and we are using all frequency bands). The likelihood for $r$ post-delensing, shown in the right panel, is nearly identical to the likelihood calculated with the fiducial 73% that has been used in section 5.

Figure 11 gives a visualization of a different metric for PICO's delensing efficacy. We show the ten-simulation average spectrum of the input lensing $B^{\rm L}$ and the averages of spectra of residual lensing maps with the QE and maximum a-posteriori iterative solvers. The spectra

– 24 –

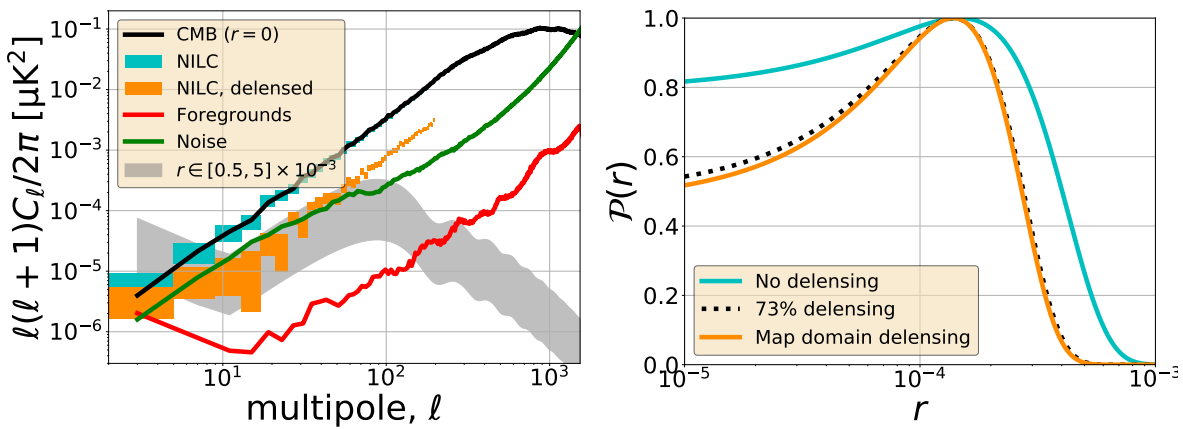

**Figure 10**. Left: Power spectra similar to those shown in figure 4 but for high-resolution $\hat{B}^{\mathrm{NILC}}$ map with model Baseline and $r = 0$. Also included is the spectrum of the delensed map (orange). Right: $r$ likelihood when $r_{\mathrm{in}} = 0$ without delensing (cyan), $r = (1.5 \pm 1.7) \times 10^{-4}$, with the approximate 73% delensing (dotted black), $r = (1.4 \pm 1.1) \times 10^{-4}$, and with map-domain delensing (orange), $r = (1.4 \pm 1.1) \times 10^{-4}$.

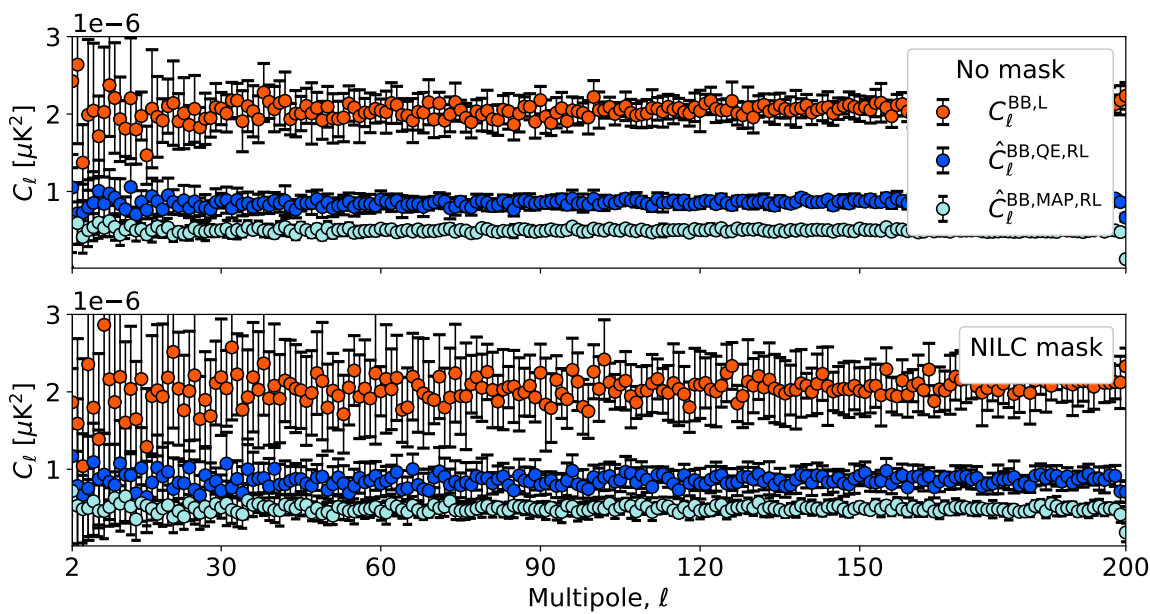

**Figure 11**. Ten-simulation average spectra of the input lensing $B^{\mathrm{L}}$ (red) and residual lensing maps $B^{\mathrm{XX,RL}}$, XX=QE (blue) or MAP (cyan), calculated over the entire sky (top) and with the NILC mask (bottom). Error bars are standard deviation over the ten simulations.

for the full and masked skies have approximately the same amplitude, with the full sky version showing a variance that is on average lower by 48% compared to the masked version, consistent with the $f_{\mathrm{sky}}$ difference. Post-delensing there is a factor of $\sim 2$ reduction in lensing power with QE and $\sim 4$ with the iterative solver. The reduction in power is essentially uniform over $2 \leq \ell \leq 200$.

| $\ell$ range | Iterative | | Quadratic Estimator | |
|---|---|---|---|---|
| | Full | Masked | Full | Masked |
| $2-200$ | $0.239 \pm 0.002$ | $0.238 \pm 0.005$ | $0.416 \pm 0.004$ | $0.416 \pm 0.005$ |
| $2-30$ | $0.25 \pm 0.01$ | $0.25 \pm 0.02$ | $0.43 \pm 0.03$ | $0.44 \pm 0.04$ |
| $30-200$ | $0.238 \pm 0.002$ | $0.238 \pm 0.005$ | $0.416 \pm 0.004$ | $0.416 \pm 0.006$ |

**Table 6**. $A_{\text{lens}}$ values obtained with the iterative and QE solvers for the full and masked sky, calculated over different ranges of angular scales. The uncertainties are the standard deviation over 10 simulations.

To quantify the delensing efficiency we calculate the residual delensing $A_{\text{lens}}$ from the ratio of the residual lensing power spectrum $C_\ell^{\text{BB,XX,RL}}$ and the input lensing $C_\ell^{\text{BB,L}}$

$$\left\langle A_{\text{lens}}^{\text{XX}} \right\rangle = \left\langle \frac{C_\ell^{\text{BB,XX,RL}}}{C_\ell^{\text{BB,L}}} \right\rangle \tag{7.4}$$

where XX is for the QE or iterative solvers, and $\langle \rangle$ denotes averaging ten values of $A_{\text{lens}}$. We calculate $A_{\text{lens}}$ values after binning the power spectra over the entire range of angular scales $2 \leq \ell \leq 200$, separately for the reionization and recombination angular scales encompassing $(\ell_{\text{reio}}, \ell_{\text{reco}}) = ([2, 30], [30, 200])$, and for the full and masked sky. The values are given in table 6.

The lowest residual lensing $A_{\text{lens}} = 24\%$ is achieved with the iterative solver when binning the entire range of angular scales, and it is consistent with $A_{\text{lens}}$ for the recombination angular scale. Within statistical uncertainty it is also consistent with the value achieved for the reionization scale $A_{\text{lens}} = 25\%$, see table 6. The uncertainty intervals given are standard deviations of the ten simulations. We use two additional methods to validate our $A_{\text{lens}}$ estimate. First, it is possible to predict the performance of the iterative solver [58, 66], and this prediction gives $A_{\text{lens}} = 0.23$ and $0.22$, for the reionization and recombination scales, respectively, which are close to the values we report. Second, the angular power spectrum domain analysis, described in section 7.1, gave $A_{\text{lens}} = 0.22$ with the PICO map noise level we assume in this paper, see table 5.

The QE solver gives $A_{\text{lens}}$ values that are nearly two times larger than obtained with the iterative solver. We find no difference between values derived for the full and masked sky. For completeness, we note that the QE result could have been further optimized by calculating the effect of $\hat{\phi}^{QE}$ on $\hat{E}^{\text{NILC,QE}}$ in a perturbative manner [71]. This approach would have led to removing at most an additional 3% of the lensing power. There is no analogous further optimization possible for the iterative solver.

## 8 Discussion and future work

The analysis that led to the constraints described in sections 5 and 6 includes several assumptions, some of which could be relaxed in future work. We discuss these below together with other possible extensions:

- We assumed spatially uniform noise. Future work could begin with time-domain simulations that include PICO's sky scan pattern and $1/f$ noise, and that would give a map with a more realistic spatial noise distribution.

- There is limited information about the behavior of foregrounds across the sky. In particular, in diffuse regions observations are limited to $\ell \lesssim 100$. While there are models that include information up to $\ell \sim 1500$, extending simulations into this regime and into even higher $\ell$ in physically realistic ways is a topic of current research.

- The analysis ignores systematic uncertainties. Preliminary analysis shows that calibration uncertainties on Planck's CMB maps give negligible errors.[2] However, we have not included calibration uncertainties with bands that cannot be calibrated on CMB maps, nor other polarimetric systematic uncertainties.

- With all sky models, the NILC component separation has been carried out over the full sky, and we made no attempts to vary NILC parameters such as the shapes of the needlet window functions. Varying these parameters or masking the Galactic plane prior to component separation might lead to smaller residual foregrounds. With model MultiLayer, masking the Galactic plane prior to component separation might reduce the bias on $r$, although this is not guaranteed because the NILC algorithm is already fairly local in pixel space. Another strategy to reduce residual foreground biases, when they occur, would be to use a more constrained version of NILC, such as cMILC [19] to enforce the minimization of the foreground variance instead of the overall variance. With this approach one deprojects the main statistical moments of the foreground emission [72] out of the reconstructed CMB map.

- We chose to apply the same conservative mask to all sky models so as to compare their forecasts from the same portion of the sky. In a further study, the mask could be optimized separately for each sky model. The choice of the mask could be informed by the small patch analysis carried out for models Baseline and MultiLayer.

- Estimates for $r$ after component separation with NILC used a Gaussian approximation for the likelihood, and an approximation to account for delensing. While for $\ell < 12$ we showed that the exact likelihood gives weaker constraints by a factor of 1.3 relative to the Gaussian approximation, the magnitude of this factor for higher $\ell$ limit is not known. Including the effect of delensing into the $r$ likelihood is also a topic of current active research.

- In this paper we took the focal plane configuration from [4]. We did not attempt to optimize the focal plane as a function of $r$ limits, neither in terms of the number of frequency bands, nor in terms of the noise level per band.

- Future work could extend to other foreground models, and to assessing the performance of parametric component separation, with Commander or otherwise, with more than the single model we included in this paper and over a larger range in $\ell$.

Model MultiLayer stands out as a model that is plausible from a physics point of view, yet posing challenges to the component separation process, even with PICO's low noise and large number of frequency bands. The spectral decorrelation of thermal dust due to multi-layer contributions along the line-of-sight makes such foreground sky more complex because of the larger dimension of the foreground subspace. Model MultiLayer was not included in the analysis of CMB-S4 et al. [21], which analyzed the expected performance of CMB-S4 over a small, ∼ 5% area of the sky, and with somewhat looser target constraints on $r$. There is therefore no information on potential biases with a narrower frequency coverage.

---

[2]z.umn.edu/picomission.

We have shown, however, that PICO's deep integration over the full sky and its high frequency bands that are only sensitive to Galactic dust, enable a powerful cross check by analyzing individual small patches of the sky and correlating the inferred $r$ determinations or limits with the variance in dust polarized intensity; see section 5.2.1. The analysis methodology we presented — quantifying $r$ constraints as a function of observed local foreground polarized intensity and searching for asymptotic behavior as a function of decreasing foreground levels — only uses data as observed by the mission, and parallels the process that is likely to be pursued by an analysis team. The fact that $r$ determinations or limits converge to the same value does not guarantee the absence of bias, but it is, however, a strong systematic cross-check. The more sky areas that give the same result, the higher the confidence there is no bias.

The small-patch analysis also suggests that constraining observations to the lowest dust polarized intensity variance regions could provide immunity from biased detections. We showed that only having the region away from the Galactic plane is not sufficient. The analysis we presented relies on data from a relatively high frequency band, which is entirely dominated by dust and may not be available for other instruments. We have not repeated the analysis with other instrument noise levels, nor with a different mix of frequency bands.

Parametric component separation with Commander has only been attempted with model Baseline. Both $r = 0$ and $r = 0.003$ are recovered with no bias. No delensing has been applied and therefore the constraints achieved cannot be compared to PICO's requirements. The constraints are nevertheless useful because they illustrate the limitations of using a Gaussian likelihood approximation, specifically when limiting the analysis to low $\ell$ multipoles. Commander uses a Blackwell-Rao approximation, which, as we show in the appendix, closely matches the exact likelihood in the case of full sky coverage and uniform noise. Using the Blackwell-Rao approximation, and when $\ell \leq 12$, Commander gives a factor of two higher upper limits compared to NILC. We note that in the NILC analysis we use $\ell \leq 200$. Assessing the differences between the exact, Blackwell-Rao, and Gaussian likelihoods over this extended $\ell$ range is computationally demanding and there are currently no quantitative estimates.

## 9 Summary

We used blind and parametric component separation techniques to assess PICO's capability to reach its $r$ determination requirements. The analysis employs realistic noise levels, only the prime mission duration without any extensions (even though the mission has no consumables), sky foreground models that match current data and are broadly accepted as plausible, and analysis techniques that have been implemented with other instruments including Planck. The analysis assumes a spatially uniform noise with white spectrum, and does not include systematic uncertainties. We initially used an angular power spectrum domain analytic calculation to find a predicted level of delensing and took the most conservative value obtained for the residual delensing factor $A_{\mathrm{lens}} = 0.27$.

For four out of the five sky models, Baseline, 2MBB, PhysDust, and MHD, and with $r_{\mathrm{in}} = 0$ the PICO experiment configuration that has 21 frequency bands gives 95% upper limits between $1.3 \times 10^{-4}$ and $2.7 \times 10^{-4}$; see table 3. These constraints could definitively rule out the simplest models of inflation that predict $r \simeq 0.001$. With the same four models, if the true value of $r$ is $r_{\mathrm{in}} = 0.003$ it would be detected with confidence levels between $18\sigma$ and $27\sigma$ after 5 yrs of the prime mission. The $r$ estimates are based on a Gaussian approximation to the likelihood. They are the strongest upper limits and detections predicted for any instrument in the foreseeable future.

Removal of the lowest or the highest frequency bands weakens the upper limits when $r_{\rm in} = 0$, and gives more biased $r$ likelihoods; see table 4 and figure 5. With models Baseline and MHD, nearly $3\sigma$ biases occur with removal of either low or high frequency bands. Although removal of the higher frequency bands results in somewhat more pronounced biases, to achieve its requirements a space mission like PICO requires bands over the entire frequency range.

With model MultiLayer the $r$ estimates are significantly biased. However, PICO's deep integration over the full sky and its high frequency bands, enables identification and mitigation of the bias. Model MultiLayer is an example for the limitation of inference using an $r$ determination over a small patch of sky and with limited frequency band data.

Parametric component separation with Commander has only been attempted with model Baseline and without delensing. Both $r=0$ and $r=0.003$ are recovered with no bias.

We presented the first application of map-domain iterative delensing, and it is the first application of iterative delensing on maps that have first undergone a component separation exercise. The delensing has been applied over the entire sky because the map for model Baseline after component separation does not indicate the need to mask any region, including along the Galactic plane; see figure 9.

The map-domain delensing gives a residual delensing factor $A_{\rm lens}= 0.24$, which would have been $A_{\rm lens}= 0.22$ had the range of multipoles $\ell < 200$ been included in the delensing process; see section 7.2.1. This level reproduces the level we calculated using the analytic approximation, which predicted $A_{\rm lens}= 0.22$ for the noise levels we assumed here; see section 7.1. The map-domain delensing confirms that the choice we made to use $A_{\rm lens}= 0.27$ is conservative.

## A Comparison of Gaussian and Blackwell-Rao likelihoods

In this appendix we compare the Gaussian approximation used in the NILC analysis and the Blackwell-Rao estimator used in the Commander analysis. We consider only ideal primordial Gaussian CMB fluctuations and instrumental noise, and ignore instrumental systematic effects, foreground subtraction, and other potential complications. Explicitly, we assume that the data $d$ may be written as $d = s + n$, where $s$ is a Gaussian distributed true CMB signal with covariance matrix $S$, and $n$ is Gaussian distributed instrumental noise with covariance matrix $N$. The sum of the two terms is Gaussian distributed with total covariance $S + N$, and the full exact likelihood is

$$\mathcal{L}_0 = \frac{\exp\left(-\frac{1}{2}d^T(S+N)^{-1}d\right)}{\sqrt{|S+N|}}. \tag{A.1}$$

The covariance $N$ is typically a deterministic quantity, given by the noise and scanning strategy of the instrument, and $S$ is typically defined by additionally assuming that the CMB is statistically isotropic. In this case $S$ is block-diagonal in the spherical harmonic domain, and $S_{lm,l'm'} = C_\ell \delta_{ll'} \delta_{mm'}$, where $C_l$ is the angular power spectrum, which is a function of the cosmological parameters. In our case, $S = S(r)$.

For full sky data with uniform noise per pixel equation (A.1) may be written in harmonic space

$$\mathcal{L}_1 = \prod_\ell \frac{\exp\left(-\frac{2\ell+1}{2}\frac{\sigma_\ell}{b_\ell^2 p_\ell^2 C_\ell + N_\ell}\right)}{(b_\ell^2 p_\ell^2 C_\ell + N_\ell)^{\frac{2\ell+1}{2}}}, \tag{A.2}$$

JCAP06(2023)034

$$\sigma_\ell \equiv \frac{1}{2\ell+1} \sum_{\ell=-m}^{m} |a_{\ell m}|^2, \tag{A.3}$$

$$N_\ell = \sigma_0^2 4\pi/N_{\text{pix}}, \tag{A.4}$$

where $\sigma_\ell$ is the observed realization-specific power spectrum, $N_\ell$ is the noise power spectrum, $\sigma_0$ is the noise per pixel, and $b_\ell$ and $p_\ell$ are the beam and pixel window transfer functions, respectively. In order to use this expression to estimate $r$, one may write [20]

$$C_\ell \equiv C_\ell(r, A_{\text{lens}}) = rC_\ell^{\text{tens}} + A_{\text{lens}} C_\ell^{\text{lens}}, \tag{A.5}$$

where the total power $C_\ell$ has contributions from the inflationary, tensor term and from the lensing term, which has a delensing factor $1 - A_{\text{lens}}$, and a residual lensing factor $A_{\text{lens}}$. When $A_{\text{lens}}=1$ there is no delensing. The nominal 0.73 delensing factor value used with NILC in section 5 leaves $A_{\text{lens}}= 0.27$ of the original $C_\ell^{\text{lens}}$ power in the map. In this expression, we implicitly assume that the net distortion effect from lensing may also be approximated as Gaussian.

The likelihood $\mathcal{L}_1$ is not valid for data with a spatially varying noise distribution or when the data do not include the full sky. Even for uniform noise and full sky, calculating $\mathcal{L}_0$ is too computationally expensive to evaluate by brute-force methods due to expensive matrix inversion and determinant operations. For these reasons many likelihood approaches have been developed that aim to approximate $\mathcal{L}_0$. One common approach is to assume that $\mathcal{L}(C_l)$ is close to a Gaussian distribution. In fact, $\mathcal{L}_1$ is an inverse Gamma distribution in $C_l$, which converges to a Gaussian for large $\ell$. For $\ell \gtrsim 30$ the Gaussian approximation is acceptable [e.g., 73].

Motivated by this observation, the NILC component separation pipeline uses the following expression for the likelihood

$$\mathcal{L}_3 \propto \exp\left(-\frac{1}{2}\sum_\ell \frac{\left(\widehat{C}_\ell^{BB,\text{NILC}} - C_\ell(r, A_{\text{lens}}) - \widehat{N}_\ell\right)^2}{\widehat{\Xi}_\ell^2}\right), \tag{A.6}$$

$$\widehat{\Xi}_\ell = \sqrt{\frac{2}{(2\ell+1)f_{\text{sky}}}}\widehat{C}_\ell^{BB,\text{NILC}}, \tag{A.7}$$

where $\hat{N}_\ell$ is an estimate of the NILC B-mode noise power spectrum after component separation, $\widehat{C}_\ell^{BB,\text{NILC}}$ is the total measured power spectrum after component separation (see section 5) and possibly delensing (see section 7.2), and $\widehat{\Xi}_\ell$ is the corresponding variance, which implicitly includes the cosmic and sample variance contributions of the primordial signal (if any), the residual lensing signal, the residual foreground signal, and the noise. When using NILC to estimate $r$, we evaluate $\mathcal{L}_3$ as a function of $r$.

Commander addresses the computational cost associated with evaluating $\mathcal{L}_0$ through Bayesian Monte Carlo sampling. With this method one first draws many Monte Carlo samples from the joint sky signal and power spectrum distribution, i.e., $P(s, C_l|d)$. At first sight, this may appear more complicated than working directly with $P(C_l|d)$. However, joint samples may be drawn through a particularly efficient Gibbs sampling algorithm [74, 75] by first sampling from $P(s|d, C_l)$ and then from $P(C_l|d, s)$, and this two-step approach is in fact computationally far cheaper than evaluating the marginal distribution directly. Each sky map sample $s^i$ represents one possible ideal CMB sky map consistent with the observed data, and the histogram of all these Monte Carlo samples converges to the true likelihood in equation (A.1).

A commonly used statistical method for speeding up the convergence rate of a Monte Carlo estimator is the Blackwell-Rao (BR) approximation [52]. Rather than building up a histogram directly from the sample set, which is subject to a high Monte Carlo uncertainty, this method instead averages the analytical probability distributions for each sample. In our case, this is equivalent to averaging $\mathcal{L}_1$ over all $s^i$. The advantage of this method is that the Monte Carlo sampler only needs to map out uncertainties due to instrumental noise and masks, but not cosmic variance; that source of uncertainty is instead handled analytically by the expression in equation (A.2).

In practice, Monte Carlo convergence can be improved even further through 'Gaussianization' [76]. In that approach, one first averages $\mathcal{L}_1$ over a large number of samples, setting $b_\ell = p_\ell = 1$ for each $\ell$. We call the resulting distribution $\mathcal{L}_\ell(C_\ell)$. This is then used to define a transformation $x_\ell(C_\ell)$ that maps $\mathcal{L}_l(C_l)$ to a perfect Gaussian by matching their cumulative distributions percentile-by-percentile. One then corrects for correlations between $\ell$s using

$$\mathcal{L}_2 = P(C_\ell | d) \approx \left( \prod_\ell \frac{\partial C_\ell}{\partial x_\ell} \right)^{-1} \exp\left( -\frac{1}{2}(x - \mu)^T C^{-1} (x - \mu) \right). \tag{A.8}$$

Here, $x = \{x_\ell(C_\ell)\}$ is the vector of Gaussianized input power spectrum coefficients, $\frac{\partial C_\ell}{\partial x_\ell}$ is the Jacobian of this Gaussianizing transformation, and the mean $\mu$ and covariance matrix $C_{\ell\ell'}$ are estimated from Monte Carlo samples of the full posterior distribution. The advantage of this approach is that only second-order correlations need to be mapped out through Monte Carlo sampling, as opposed to all possible $N$-point correlations, and this greatly speeds up the overall convergence rate. The disadvantage is that the expression is no longer an exact representation of $\mathcal{L}_1$, and it can break down for data that cover only small sky fractions and that have strong correlations. However, for $f_{\text{sky}} \gtrsim 0.70$, as is typical for CMB satellite experiments, it is an excellent approximation [76].

We quantitatively compare the Gaussian and the BR approximations to the exact likelihood $\mathcal{L}_0$. We generate ten CMB full sky simulations that all have $r = 0$ and a noise level of $\sigma_0 = 0.3\,\mu\text{K/pixel}$ at $N_{\text{side}} = 512$, set by the estimated NILC noise level shown in figure 4. We omit delensing to simplify the comparison. We compare the $r_{95\%}$ upper limits obtained with $\mathcal{L}_1$, $\mathcal{L}_2$, and $\mathcal{L}_3$ using $2 \le \ell \le 12$. The resulting posterior distributions are thus not representative of the full PICO sensitivity, which also includes higher multipoles, but are only useful for comparing the three likelihood approximations. Figure 12 shows the $r$ likelihoods for two of these simulations, one for which the likelihood peaks at zero (left panel) and another for which a noise fluctuation gives a non-zero value for the peak of the likelihood (right panel). In both cases the BR approach gives a larger $r_{95\%}$, although this is more pronounced when the peak has a non-zero value. In both cases the BR likelihood $\mathcal{L}_2$ agrees quite closely with the exact likelihood $\mathcal{L}_0$, and is in better agreement than the Gaussian approximation $\mathcal{L}_3$. This suggests that $\mathcal{L}_2$ might also be closer to $\mathcal{L}_0$ when only part of the sky is used.

The distribution of $r_{95\%}$ for the ten simulations are given in figure 13. On average the BR likelihood predicts upper limits that are a factor of 1.3 times larger than the Gaussian likelihood. However, this may be split into two classes, those that peak at zero and those that peak at a non-zero value. Examination of the likelihoods indicates that the Gaussian approximation performs well for the former class, while for the latter, the BR $r_{95\%}$ value may be up to nearly two times larger than the Gaussian, as happens in realizations 5 and 8.

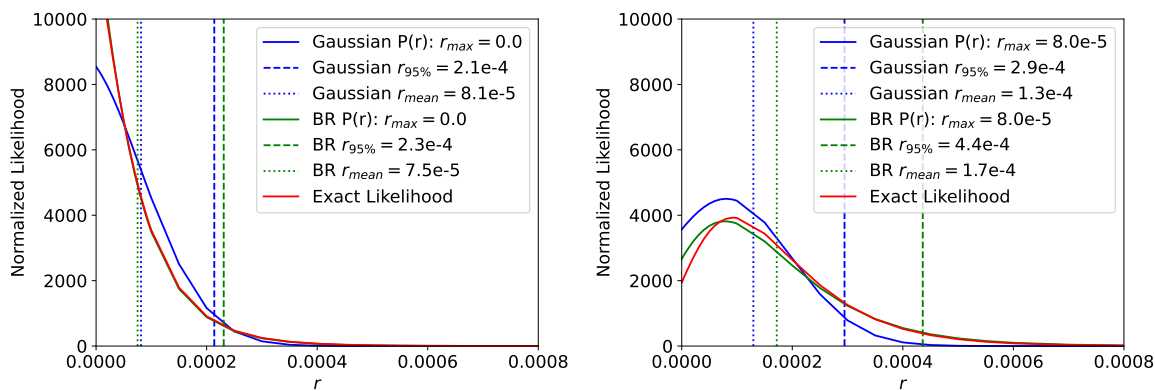

**Figure 12**. Two simulations with $r = 0$ (left and right) comparing the likelihood curves calculated using the exact (red), Gaussian (blue), and BR (green) likelihoods. In all cases $r$ is estimated for $2 \leq \ell \leq 12$.

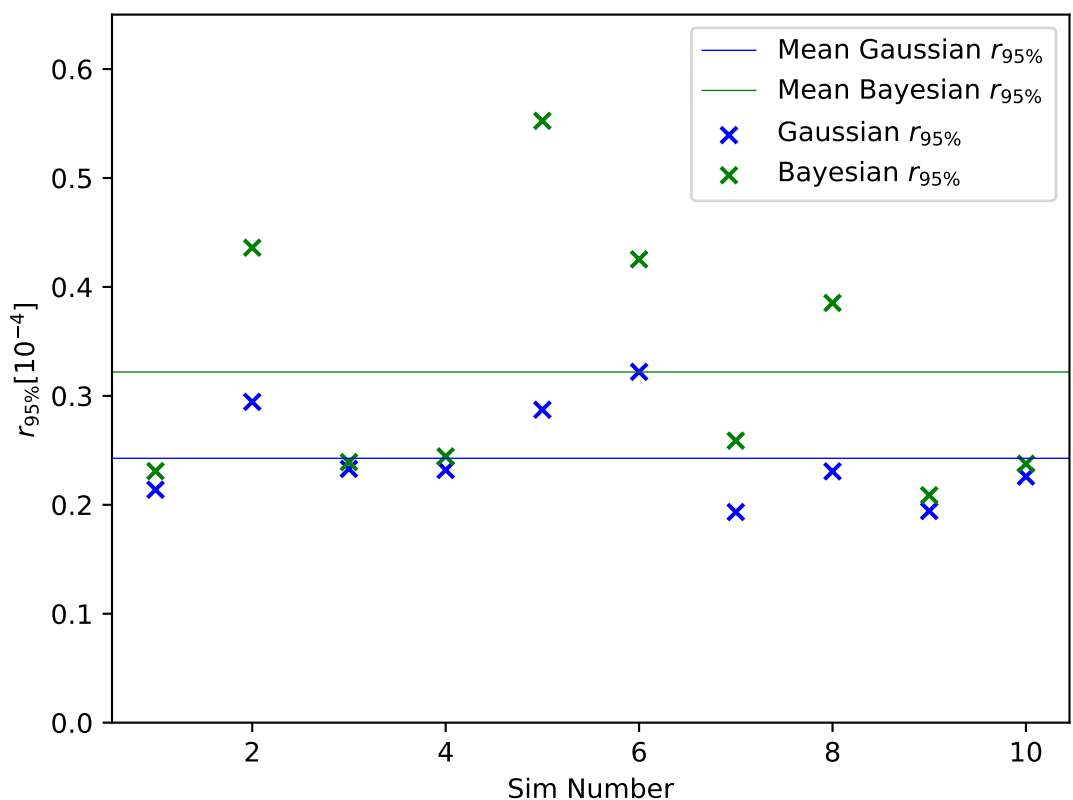

**Figure 13**. $r_{95\%}$ limits for ten $r = 0$ CMB and noise simulations, for the Gaussian (blue) and BR (green) likelihoods. The averages of the two sets ($r = 2.4 \times 10^{-4}$ for Gaussian and $3.2 \times 10^{-4}$ for BR) are plotted as the horizontal lines.

## Acknowledgments

This research used resources of the National Energy Research Scientific Computing Center, which is supported by the Office of Science of the U.S. Department of Energy under Contract No. DE-AC02-05CH11231. MR would like to thank the Spanish Agencia Estatal de Investigación (AEI, MICIU) for the financial support provided under the project with reference PID2019-110610RB-C21.